\documentclass[runningheads,a4paper]{llncs}

\usepackage{amssymb}
\usepackage{graphicx}

\usepackage[cmex10]{amsmath}

\usepackage{subfig}

\usepackage{url}
\newcommand{\keywords}[1]{\par\addvspace\baselineskip
\noindent\keywordname\enspace\ignorespaces#1}

\begin{document}

\mainmatter

\title{The Inner Structure of Time-Dependent Signals}

\titlerunning{The Inner Structure of Time-Dependent Signals}

\author{David N. Levin%
}

\authorrunning{David N. Levin}

\institute{Dept. of Radiology, University of Chicago,\\
1310 N. Ritchie Ct., Unit 26 AD, Chicago, IL 60610\\
Email: d-levin@uchicago.edu\\
\url{http://radiology.uchicago.edu/directory/david-n-levin}}

\toctitle{}
\tocauthor{}
\maketitle

\begin{abstract}
This paper shows how a time series of measurements of an evolving system can be processed to create an ''inner" time series that is unaffected by any instantaneous invertible, possibly nonlinear transformation of the measurements. An inner time series contains information that does not depend on the nature of the sensors, which the observer chose to monitor the system. Instead, it encodes information that is intrinsic to the evolution of the observed system. Because of its sensor-independence, an inner time series may produce fewer false negatives when it is used to detect events in the presence of sensor drift. Furthermore, if the observed physical system is comprised of non-interacting subsystems, its inner time series is separable; i.e., it consists of a collection of time series, each one being the inner time series of an isolated subsystem. Because of this property, an inner time series can be used to detect a specific behavior of one of the independent subsystems without using blind source separation to disentangle that subsystem from the others. The method is illustrated by applying it to: 1) an analytic example; 2) the audio waveform of one speaker; 3) video images from a moving camera; 4) mixtures of audio waveforms of two speakers.   
\end{abstract}

\section{Introduction}
\label{introduction}

Consider a physical system that is being observed with a set of sensors. The time series of raw sensor measurements contains information about the evolution of the system of interest, mixed with information about the nature of the sensors. For example, video pictures contain information about the evolution of the scene of interest, but they are also influenced by sensor-dependent factors such as the position, angular orientation, field of view, and spectral response of the camera. Likewise, audio measurements may describe the evolution of an acoustic source, but they are also influenced by extrinsic factors such as the positions and frequency responses of the microphones. Calibration procedures can be used to transform measurements created with one set of sensors so that they can be compared to measurements made with a different set of sensors (\cite{Elbert},\cite{Morain},\cite{Bottomley}). However, there are situations in which it is inconvenient, awkward, or impossible to calibrate a measurement apparatus. For example: 1) the calibration procedure may take too much time; 2) the calibration process may interfere with the evolution of the system being observed; 3) the observer may not have access to the measuring device (e.g., because it is at a remote location).

This paper describes how a time series of measurements can be processed to derive a purely sensor-independent description of the evolution of the underlying physical system. Specifically, consider an evolving physical system with $N$ degrees of freedom ($N \geq 1$), and suppose that it is being observed by $N$ sensors, whose output is denoted by $x(t)$ ($x_{k}(t) \mbox{ for } k = 1, \ldots ,N$). For simplicity, assume that the sensors' output is invertibly related to the system states. In other words, assume that the sensor measurements represent the system's state in a coordinate system defined by the nature of the sensors. Section \ref{conclusion} describes how measurements can be chosen to have this invertibility property. Now, suppose that the same system is also being observed by another set of sensors, whose output, $x'(t)$, is invertibly related to the system states and, therefore, is invertibly related to $x(t)$. For example, $x(t)$ and $x'(t)$ could be the outputs of calibrated and uncalibrated sensors, respectively, as they simultaneously observe the same system. Or, they could be the outputs of sensors that detect different types of energy (e.g., infrared light vs ultraviolet light). Under these conditions, we show how to process $x(t)$ in order to derive an ''inner" time series, $w(t)$ ($w_{k}(t) \mbox{ for } k = 1, \ldots ,N$). We then demonstrate that the same inner time series will result if the other set of sensor outputs, $x'(t)$, is subjected to the same procedure. Because of its sensor-independence, an inner time series may produce fewer false negatives when it is used to detect events in the presence of sensor drift. In mathematical terms, $x(t)$ and $x'(t)$ represent the evolving system's state in different coordinate systems on state space, and the inner time series is a coordinate-system-independent description of the system's velocity in state space.

To derive this sensor-independent time series, the time series of sensor measurements, $x(t)$, is statistically processed in order to construct $N$ local vectors at each point in state space. The system's path through state space can then be described by a succession of small displacement vectors, each of which is a weighted superposition of the local vectors. The inner time series is comprised of these time-dependent weights, $w(t)$, which are coordinate-system-independent and, therefore, sensor-independent. Thus, any two observers will describe the system's evolution with the same inner time series, even though they utilize different sensors to monitor the system. Essentially, an inner time series is a ''canonical" form of a measurement time series, created by normalizing the measurements with respect to their own statistical properties. No matter what linear or nonlinear transformation has been applied to a sequence of measurements, its canonical form (i.e., its inner time series) is the same. An inner time series is roughly analogous to the principal components of a data set, which represent the data in the same ''canonical" way, no matter what rotation and/or translation has been applied to them.

There are many ways of using a time series of measurements to define local vectors on the system's state space, and each of these methods can be used to create a sensor-independent description of the system's evolution. However, the local vectors described in this paper have an unusually attractive property: namely, they produce \textit{separable} sensor-independent descriptions of systems that are composed of non-interacting subsystems. Specifically, consider a system that is composed of two statistically independent subsystems, and suppose that the raw measurements of it are linear or nonlinear mixtures of the state variables of its non-interacting subsystems. It can be shown that each component of the inner time series of the composite system is also a component of the inner time series of an isolated subsystem. In other words, each component of the inner time series of the composite system is a stream of information about just \textit{one} of the subsystems, even though it may have been derived from measurements sensitive to several subsystems. Because of this property, an inner time series can be used to detect a specific behavior of one subsystem, which is evolving in the presence of other subsystems. In contrast to blind source separation procedures (\cite{Comon Jutten}, \cite{Jutten}, \cite{Almeida}), this is done without finding the mixing function, which relates the raw measurements of the composite system to the states of its subsystems.

Reference \cite{Levin ci-JAP} describes a different way of creating sensor-independent representations of evolving systems. First, the second-order correlations of the system's local velocity distributions are used to define a Riemannian metric and affine connection on the manifold of measurements. Then, each incremental displacement along the system's path through state space is described as a superposition of reference vectors, parallel transferred from the beginning of the path. Such a description will be coordinate-system-independent (and, therefore, sensor-independent), if it includes a coordinate-system-independent way of identifying the reference vectors at the initial point of each path of interest. In contrast, the method proposed in the current paper does not require reference vectors; instead it utilizes local vectors that are properties of the local velocity distributions of the system's past trajectory. Furthermore, the methodology in \cite{Levin ci-JAP} does not provide a simple description of composite systems. In contrast, the method proposed here always creates a sensor-independent description of a composite system, consisting of a collection of the sensor-independent descriptions of the independent subsystems.

The next section describes the procedure for computing an inner time series from a time series of raw measurements. It also demonstrates that the inner time series of a composite system consists of a collection of the inner time series of its constituent parts. Section \ref{examples} illustrates the method by applying it to: 1) an analytic example; 2) the audio waveform of one speaker; 3) video images from a moving camera; 4) mixtures of audio waveforms of two speakers. The last section discusses the implications of this approach.

\section{Method}
\label{method}

The following subsection outlines how a time series of sensor measurements can be processed in order to derive local vectors at each point in the state space of the observed system. This procedure is only presented in outline form here, because detailed descriptions can be found in \cite{Levin arXiv} and \cite{Levin LVA-ICA}. It is then shown how these vectors can be used to create an inner description of the system's path through state space. In the second subsection, the system is assumed to be composed of two statistically independent subsystems. It is shown that the inner time series of the composite system is a simple collection of the inner time series of its subsystems.

\subsection{Derivation of inner time series}
\label{derivation} 

 The first step is to construct second-order and fourth-order local correlations of the data's velocity ($\dot{x}$)
\begin{equation}
\label{C2 definition}
C_{kl}(x) = \, \langle (\dot{x}_k-\bar{\dot{x}}_k) (\dot{x}_l-\bar{\dot{x}}_l) \rangle_{x} 
\end{equation}
\begin{equation}
\label{C4 definition}
C_{klmn}(x) = \, \langle (\dot{x}_k-\bar{\dot{x}}_k) (\dot{x}_l-\bar{\dot{x}}_l) 
(\dot{x}_m-\bar{\dot{x}}_m) (\dot{x}_n-\bar{\dot{x}}_n) \rangle_{x}
\end{equation}
where $\bar{\dot{x}} = \langle \dot{x} \rangle_x$, where the bracket denotes the time average over the trajectory's segments in a small neighborhood of $x$, and where all subscripts are integers between $1$ and $N$ with $N \geq 1$.

Next, let $M(x)$ be any local $N \times N$ matrix, and use it to define $M \mbox{-transformed}$ velocity correlations, $I_{kl}$ and $I_{klmn}$
\begin{equation}
\label{I2 definition}
I_{kl}(x) = \sum_{1 \leq k', \, l' \leq N} M_{kk'}(x) M_{ll'}(x) C_{k'l'}(x) ,
\end{equation}
\begin{equation}
\label{I4 definition}
I_{klmn}(x) = \sum_{1 \leq k', \, l', \, m', \, n' \leq N} M_{kk'}(x) M_{ll'}(x) 
M_{mm'}(x) M_{nn'}(x) C_{k'l'm'n'}(x) .
\end{equation}
Because $C_{kl}(x)$ is generically positive definite at any point $x$, it is almost always possible to find a particular form of $M(x)$ that satisfies
\begin{equation}
\label{M definition 1}
I_{kl}(x) = \delta_{kl}
\end{equation}
\begin{equation}
\label{M definition 2}
\sum_{1 \leq m \leq N} I_{klmm}(x) = D_{kl}(x) , 
\end{equation}
where $D(x)$ is a diagonal $N \times N$ matrix (\cite{Levin arXiv}, \cite{Levin LVA-ICA}). As long as $D$ is not degenerate, $M(x)$ is unique, up to arbitrary \textit{local} permutations and/or reflections. In almost all applications of interest, the velocity correlations will be continuous functions of $x$. Therefore, in any neighborhood of state space, there will always be a continuous solution for $M(x)$, and this solution is unique, up to arbitrary \textit{global} permutations and/or reflections. 

In any other coordinate system $x'$, the most general solution for $M'$ is given by
\begin{equation}
\label{M'}
M'_{kl}(x') = \sum_{1 \leq m, \, n \leq N} P_{km} M_{mn}(x) \frac{\partial x_n}{ \partial x'_l} ,
\end{equation}
where $M$ is a matrix that satisfies (\ref{M definition 1}) and (\ref{M definition 2}) in the $x$ coordinate system and where $P$ is a product of permutation, reflection, and identity matrices (\cite{Levin arXiv}, \cite{Levin LVA-ICA}). By construction, $M$ is not singular.

Notice that (\ref{M'}) shows that the rows of $M$ transform as local covariant vectors, up to a global permutation and/or reflection. Likewise, the same equation implies that the columns of $M^{-1}$ transform as local contravariant vectors (denoted as $V_{(i)}(x) \mbox{ for } i = 1, \ldots N$), up to a global permutation and/or reflection. Because these vectors are linearly independent, the measurement velocity at each time ($\dot{x}(t)$) can be represented by a weighted superposition of them
\begin{equation}
\label{xDot rep}
\dot{x}(t) = \sum_{1 \leq i \leq N} w_{i}(t) V_{(i)}(x)  ,
\end{equation}
where $w_{i}$ are time-dependent weights. Because $\dot{x}$ and $V_{(i)}$ transform as contravariant vectors (except for a possible global permutation and/or reflection), the weights $w_{i}$ must transform as scalars or invariants; i.e., they are independent of the coordinate system in which they are computed (except for a possible permutation and/or reflection). Therefore, the time-dependent weights, $w_{i}(t)$, provide an inner (coordinate-system-independent) description of the system's velocity in state space. Two observers, who use different sensors (and, therefore, different state space coordinate systems), will derive the same inner time series, except for a possible global permutation and/or reflection.

This equation can be integrated over the time interval $[t_{0},t]$ to give an expression for the system's state during that time interval
\begin{equation}
\label{x rep}
x(t) = x(t_0) + \int_{t_0}^{t} \sum_{1 \leq i \leq N} w_{i}(t) V_{(i)}[x(t)] dt  ,
\end{equation}
This is an integral equation for constructing $x(t)$ on the interval $[t_{0}, t]$ from the weight time series, $w_{i}(t)$, on the same time interval. Note that, given a set of local vectors, there is a many-to-one correspondence between the set of measurement time series and corresponding inner time series. Specifically, (\ref{xDot rep}) shows that each measurement time series maps onto just one weight time series. However, as shown by (\ref{x rep}), one weight time series maps onto multiple time series of sensor measurements, differing by the choice of the initial point, $x(t_0)$. It should also be mentioned that it may be difficult to use this equation to numerically compute the measurement time series, corresponding to a given weight time series, because errors will tend to accumulate as one integrates the right side.

\subsection{Inner time series of composite systems}
\label{composite systems}

Now, consider the special case in which the observed system is composite (or separable) in the sense that it consists of two statistically independent subsystems. Specifically, assume that there is a state space coordinate system, $s$, in which the state components, $s_{k}(t) \mbox{ for } k = 1, \ldots ,N$, can be partitioned into two groups, $s_{(1)} = (s_{k} \mbox{ for } k = 1, \ldots ,N_1)$ and $s_{(2)} = (s_{k} \mbox{ for } k = N_{1} + 1, \ldots ,N)$, that are statistically independent in the following sense (\cite{Levin arXiv}, \cite{Levin LVA-ICA}). Let $\rho_S(s,\dot{s})$ be the PDF in $(s,\dot{s}) \mbox{-space}$. Namely, $\rho_S(s,\dot{s}) ds d\dot{s}$ is the fraction of total time that the location and velocity of $s(t)$ are within the volume element $ds d\dot{s}$ at location $(s,\dot{s})$. The subsystem state variables, $s_{(1)}$ and $s_{(2)}$, are assumed to be statistically independent in the sense that the density function of the system variable is the product of the density functions of the two subsystem variables; i.e.,
\begin{equation}
\label{phase space factorization}
\rho_S(s,\dot{s}) = \prod_{a=1,2}{\rho_{a}(s_{(a)},\dot{s}_{(a)})} .
\end{equation}
This separability criterion in $(s,\dot{s}) \mbox{-space}$ is stronger than the conventional formulation in $s \mbox{-space}$, and references \cite{Levin arXiv} and \cite{Levin LVA-ICA} argue that this makes it preferable to the conventional criterion. In the following paragraphs, it is shown that, if the data are separable in the above sense, the components of the inner time series of the composite system can be partitioned into two groups, each of which provides an inner description of one of the subsystems. Although these results are demonstrated here for systems with two independent subsystems, they can be easily generalized to systems with any number of subsystems.

To show this, the first step is to transform (\ref{xDot rep}) into the $s$ coordinate system, by multiplying each side by $ds/dx$. Because the $V_{(i)}$ transform as contravariant vectors (up to a possible permutation and/or reflection), it follows that
\begin{equation}
\label{sDot rep}
\dot{s}(t) = \sum_{1 \leq i, \, j \leq N} w_{i}(t) P_{ij} V_{S(j)}  ,
\end{equation}
where $V_{S(j)}$ is $V_{(j)}$ in the $s$ coordinate system and $P$ is a possible permutation and/or reflection. By definition, the $V_{S(i)}$ are the local vectors, which are derived from the local distribution of $\dot{s}$ in the same way that the $V_{(i)}$ were derived from the local distribution of $\dot{x}$. Specifically, $V_{S(i)}$ is the $i^{th}$ column of $M^{-1}_S$, where $M_S$ is the $M$ matrix that is derived from the second- and fourth-order velocity correlations in the $s$ coordinate system.

The next step is to show that the matrix $M_S$ has a simple block-diagonal form. In particular, \cite{Levin arXiv} and \cite{Levin LVA-ICA} show that $M_S$ is given by
\begin{equation}
\label{block-diagonal MS}
M_S(s) = \left( \begin{array}{ccc}  
   M_{S1}(s_{(1)}) & 0 & \\
   0 & M_{S2}(s_{(2)}) &
   \end{array} \right) .
\end{equation}
where each submatrix, $M_{Sa}$ for $a = 1,2$, satisfies (\ref{M definition 1}) and (\ref{M definition 2}) for correlations between components of $s_{(a)}$. Observe that each vector $V_{S(i)}$ vanishes except where it passes through one of the blocks of $M^{-1}_S$. Therefore, equation (\ref{sDot rep}) is equivalent to a pair of equations, which are formed by projecting it onto each block corresponding to a subsystem state variable. For example, projecting both sides of (\ref{sDot rep}) onto block $a$ gives the result
\begin{equation}
\label{s(a)Dot rep}
\dot{s}_{(a)}(t) = \sum_{\substack{1 \leq i \leq N \\ j \, \in \, block \, a}} w_{i}(t) P_{ij} V_{S(ja)}  .
\end{equation}
Here, $V_{S(ja)}$ is the projection of $V_{S(j)}$ onto block $a$; i.e., it is the column of $M_{Sa}^{-1}$ that coincides with column $j$ of $M^{-1}_S$, as it passes through block $a$. This means that the vectors, $V_{S(ja)}$ for $j \in block \, a$, are the local vectors on the $s_{(a)}$ manifold, which are derived from the local distribution of $\dot{s}_{(a)}$ in the same way that the $V_{(i)}$ were derived from the local distribution of $\dot{x}$. Notice that each time-dependent weight, $w_{i}(t)$, describes the evolution of \emph{just one} subsystem. In other words, the weights do not contain a mixture of information about the evolution of the two subsystems. This is true despite the fact that they can be derived from raw measurements that may be complicated unknown mixtures of the state variables of both subsystems. 

Next, define group $1$ (group $2$) to be the set of weights appearing in the expression
\begin{equation} 
\sum_{1 \leq i \leq N} w_{i} P_{ij}
\end{equation}
for $j \in block \, 1$ (for $j \in block \, 2$). Equation (\ref{s(a)Dot rep}) shows that the weights in group $1$ (group $2$) comprise a sensor-independent description of the velocity of subsystem $1$ (subsystem $2$). Equation (\ref{s(a)Dot rep}) also suggests that the weights in group 1 must be statistically independent of the weights in group 2. Specifically, (\ref{s(a)Dot rep}) implies that the weights in each group can be computed from: 1) the time course of the state variable of the corresponding subsystem; 2) the local vectors of the corresponding subsystem, which themselves are constructed from the time course of the state variable of the corresponding subsystem. Because the weights in group 1 and group 2 are derived from $s_{1}(t)$ and $s_{2}(t)$, respectively, and because the latter are statistically independent, it is likely that the former are also statistically independent.

\section{Analytic and Experimental Examples}
\label{examples}

In this section, the methodology of Section \ref{method} is illustrated by applying it to: 1) an analytic example (namely, a time series equal to a sine wave); 2) the audio waveform of a single speaker; 3) video data from a camera moving with two degrees of freedom; 4) nonlinear mixtures of the waveforms of two speakers. 

\subsection{Analytic example: a sine wave}
\label{analytic example}

In this subsection, the proposed methodology is applied to a measurement time series, simulated by a sine wave. Its inner time series is derived analytically, before and after it is transformed by an arbitrary monotonic function. The transformed data, which simulate the output of a second sensor, are shown to have the same inner time series as the untransformed data from the first sensor.

Suppose the measured sensor signal is
\begin{equation}
\label{x analytic}
x(t) = a \sin(t)
\end{equation}
where $a$ is any real number and $- \infty \leq t \leq \infty$. Because of the periodicity of the signal, the local second-order velocity correlation can be shown to be
\begin{equation} 
C_{11}(x) = a^{2} - x^{2} .
\end{equation}
The $1 \times 1$ ``matrix", $M$, is
\begin{equation}
\label{M analytic} 
M_{11}(x) = \pm 1/ \sqrt{a^{2} - x^{2}} ,
\end{equation}
and the one-component local vector, $V_{(1)}(x)$, is
\begin{equation} 
\label{V analytic}
V_{(1)1}(x) = \pm \sqrt{a^{2} - x^{2}} .
\end{equation}
Either sign can be chosen in (\ref{M analytic}) and (\ref{V analytic}) because $M$ is only determined up to a global reflection. Substituting (\ref{x analytic}) and (\ref{V analytic}) into (\ref{xDot rep}) shows that the weight time series is
\begin{equation} 
\label{w analytic}
w_{1}(t) = \pm sgn \left[a \cos(t) \right] .
\end{equation}
Thus, for this simple periodic signal, the inner time series is the sign of the signal's time derivative. As shown in the following subsections, a much larger amount of information is contained in the inner time series of more complex one-component signals. 

The sensor-independence (or coordinate-system-independence) of the inner time series can be demonstrated explicitly by computing it from measurements that have been transformed by a monotonic function, $f(x)$, which simulates the relative response of a different sensor. Specifically, consider the transformed measurements given by
\begin{equation}
\label{x' analytic}
x'(t) = f\left[a \sin(t)\right] ,
\end{equation}
where $f$ is monotonic. The local second-order correlation of the velocity of these measurements is
\begin{equation}
\label{C' analytic}
C'_{11}(x') = \left[ \frac{df}{dx} a \cos(t_{x'}) \right]^2 ,
\end{equation}
where $df/dx$ is evaluated at $x = a \sin(t_{x'})$ and where $t_{x'}$ is any solution of $f \left[a \sin(t_{x'}) \right] = x'$. Because the measurements have just one component, the $1 \times 1$ ''matrix" $M'$ is equal to
\begin{equation}
\label{M' analytic}
M'_{11}(x') = \pm 1 / \sqrt{C'_{11}(x')} ,
\end{equation}
and the local vector is
\begin{equation}
\label{V' analytic}
V'_{(1)1}(x') = \pm \sqrt{C'_{11}(x')} .
\end{equation}
Substituting (\ref{x' analytic}) and (\ref{V' analytic}) into (\ref{xDot rep}) shows that the weight function is
\begin{equation} 
\label{w analytic}
w'_{1}(t) = \pm sgn \left[a \cos(t) \right] = w_{1}(t) .
\end{equation}
Thus, the transformed and untransformed measurements ((\ref{x' analytic}) and (\ref{x analytic})) have the same inner time series (up to a reflection), This shows that the weights are sensor-independent (and coordinate- system-independent), a fact that was proved in Section \ref{method}.

\subsection{The audio signal of a single speaker}
\label{audio signal}

In this subsection, the proposed method is applied to the audio waveform of a single speaker, before and after it has been transformed by a nonlinear monotonic function, which simulates the relative response of another sensor. The inner time series of the untransformed and transformed signals are shown to be almost the same.

\begin{figure}
\centering
\subfloat{\includegraphics[width=0.35\textwidth]{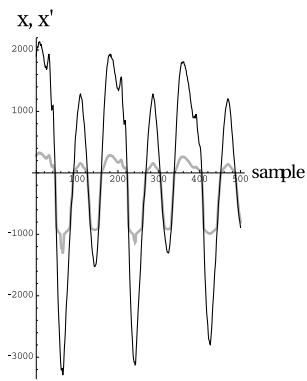}%
}
\caption{ The thin black line shows a 31.25 ms excerpt of $x(t)$, the audio waveform of a speaker. The thick gray line shows $x'(t)$, the same waveform, after it has been transformed by the monotonic nonlinear transformation shown in Figure \ref{fig_xPrimeOfX}.}
\label{fig_xXPrimeOfT}
\end{figure}

\begin{figure}
\centering
\subfloat[]{\includegraphics[width=0.35\textwidth]{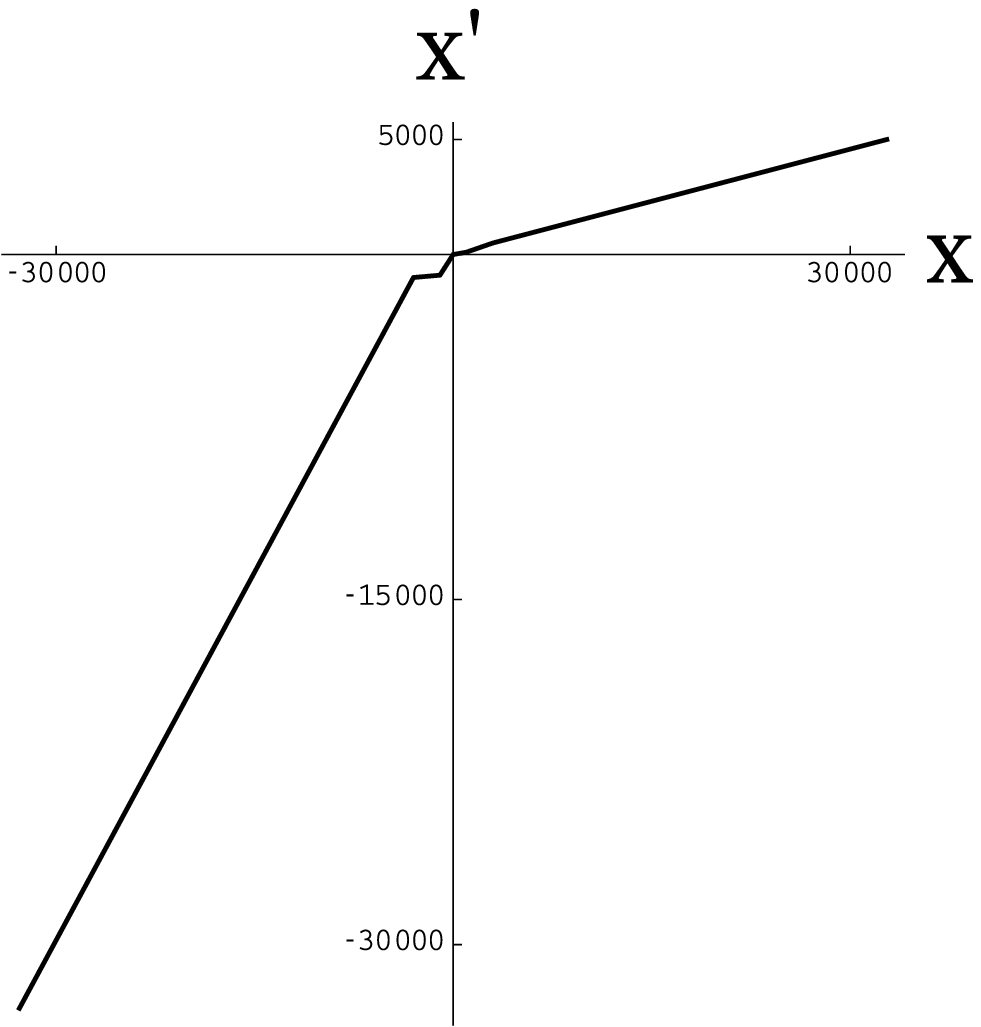}%
\label{fig_xPrimeOfXGlobal}}
\subfloat[]{\includegraphics[width=0.35\textwidth]{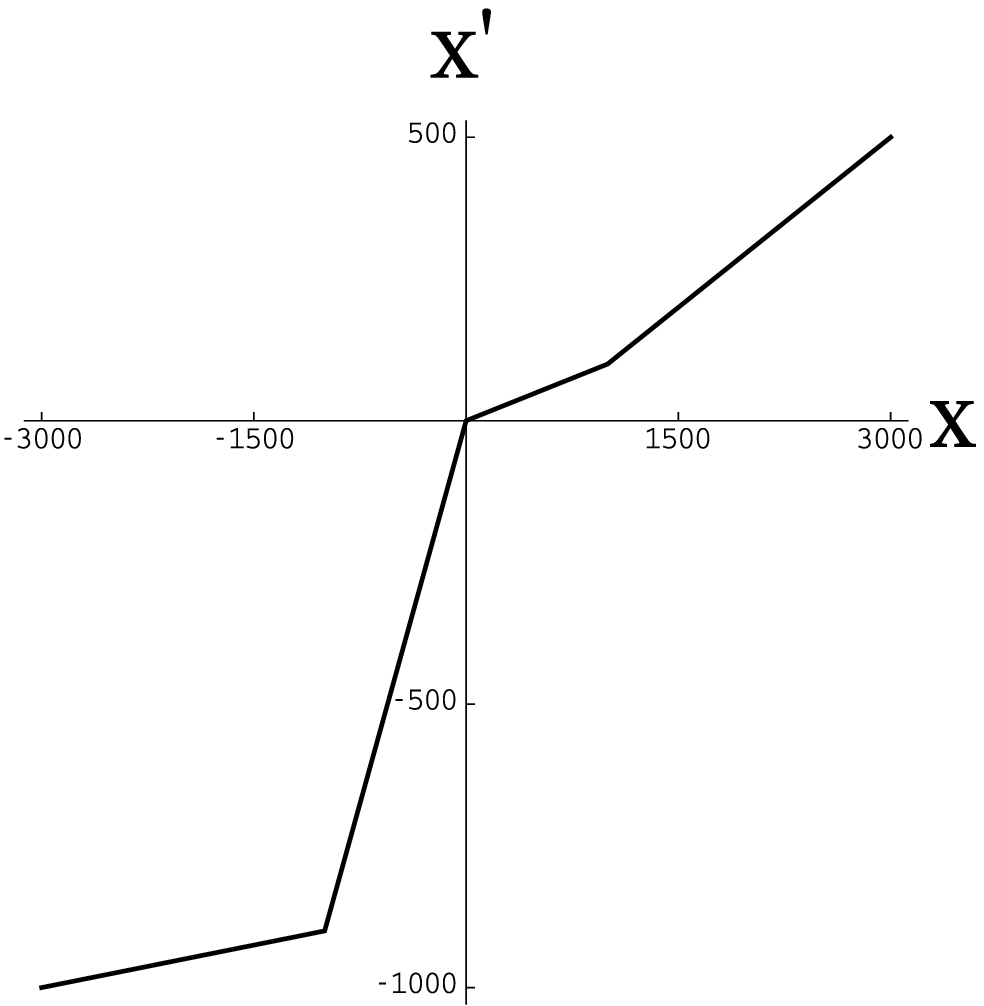}%
\label{fig_xPrimeOfXLocal}}
\caption{The left panel shows the monotonic nonlinear transformation, $x'(x)$, which was applied to the sensor measurements, $x(t)$, in order to create $x'(t)$ (Figure \ref{fig_xXPrimeOfT}). The latter time series simulates the output of a different sensor. The right panel is a magnified view of the central portion of the left panel.}
\label{fig_xPrimeOfX}
\end{figure}

The male speaker's audio waveform, $x(t)$, was a 31.25 s excerpt from an audio book recording. This waveform was sampled 16,000 times per second with two bytes of depth. The thin black line in Figure \ref{fig_xXPrimeOfT} shows the speaker's waveform during a short (31.25 ms) interval. The thick gray line in Figure \ref{fig_xXPrimeOfT}, $x'(t)$, simulates the output of another sensor, which is related to $x(t)$ by the monotonic nonlinear transformation in Figure \ref{fig_xPrimeOfX}.

\begin{figure}
\centering
\subfloat[]{\includegraphics[width=0.35\textwidth]{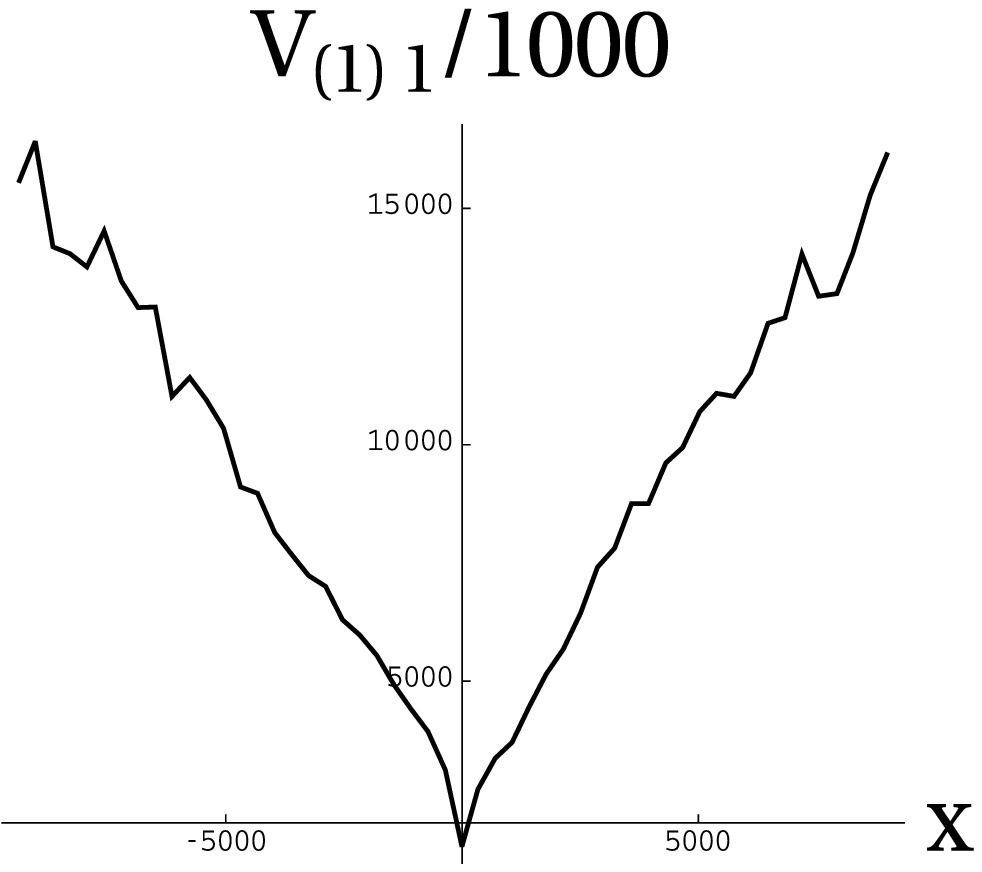}%
\label{fig_VOfX}}
\subfloat[]{\includegraphics[width=0.35\textwidth]{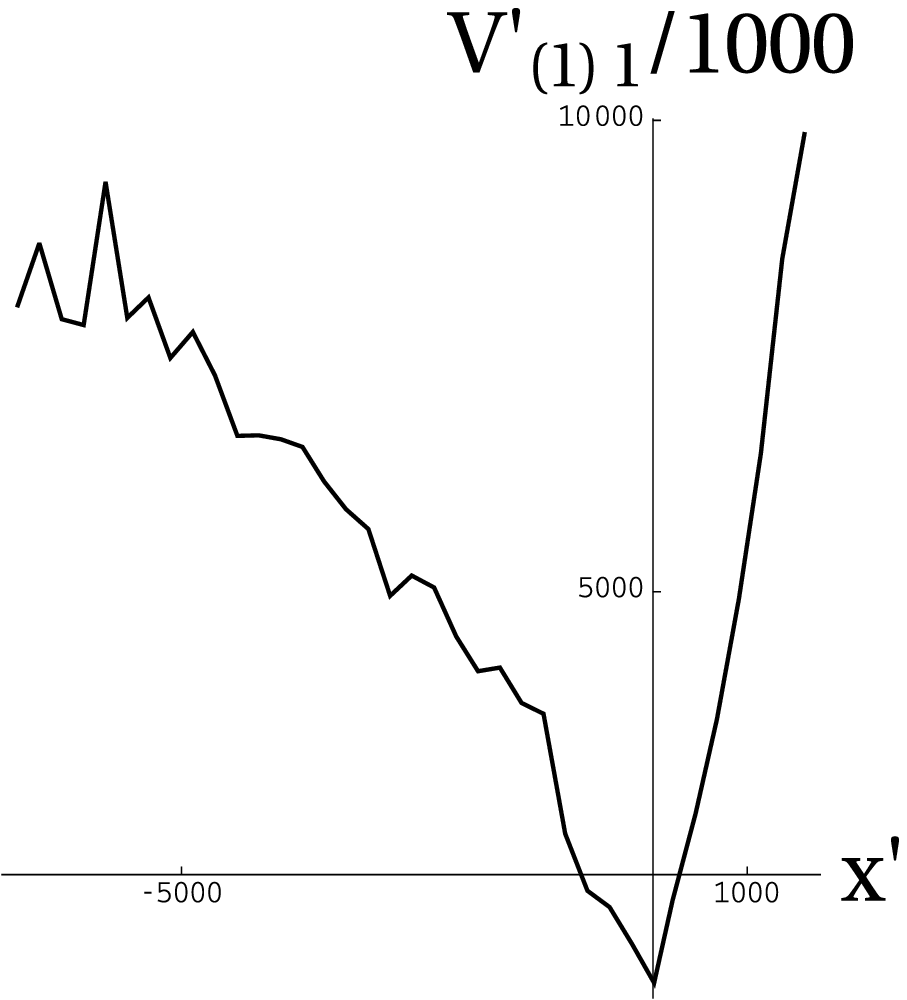}%
\label{fig_VPrimeOfXPrime}}
\caption{The left and right panels show the local vectors, $V_{(1)}(x)$ and $V'_{(1)}(x')$, which were derived from 500,000 samples of $x(t)$ and $x'(t)$, respectively.}
\label{fig_VOfXVPrimeOfXPrime}
\end{figure}

The technique in Subsection \ref{derivation} was applied to 500,000 samples of $x(t)$ and $x'(t)$, in order to derive the one-component vectors, $V_{(1)}(x)$ and $V'_{(1)}(x')$, in an array of 128 bins on the $x$ and $x'$ manifolds, respectively. These vectors are displayed in Figure \ref{fig_VOfXVPrimeOfXPrime}.

\begin{figure}
\centering
\subfloat{\includegraphics[width=0.35\textwidth]{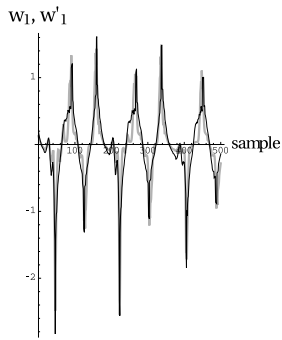}%
}
\caption{The thin black line and thick gray line show the inner time series, $w_{1}(t)$ and $w'_{1}(t)$, respectively, during the 31.25 ms time interval depicted in Figure \ref{fig_xXPrimeOfT}.}
\label{fig_wWPrimeOfT}
\end{figure}

Then, these vectors and equation (\ref{xDot rep}) were used to compute the inner time series, $w_{1}(t)$ and $w'_{1}(t)$, corresponding to the two measurement time series, $x(t)$ and $x'(t)$, respectively. The resulting time series of weights are shown in Figure \ref{fig_wWPrimeOfT}. Notice that the two inner time series are almost the same, despite the fact that they were derived from sensor measurements, which differed by a nonlinear transformation. This demonstrates the sensor-independence of the weights, a property that was proved in general in Subsection \ref{derivation}. 

When either inner time series was played as an audio file, it sounded like a completely intelligible version of the original audio waveform, $x(t)$. No semantic information was lost, although the prosody of the signal may have been modified. Therefore, in this experiment, almost all of the signal's information content was preserved by the process of deriving its inner time series.

\subsection{Video data from a moving camera}
\label{video signal}

In this subsection, the procedure in Subsection \ref{derivation} is used to derive the inner time series of a sequence of video images, recorded by a camera moving in an office. We also computed the inner time series of the same image sequence, after each image was subjected to a nonlinear transformation, thereby simulating the output of a different sensor (i.e., a different video camera). The two inner time series were almost the same, despite the fact that they were derived from the outputs of dramatically different sensors.

\begin{figure}
\begin{center}$
\begin{array}{cccccccccc}
\includegraphics[width=0.1\textwidth]{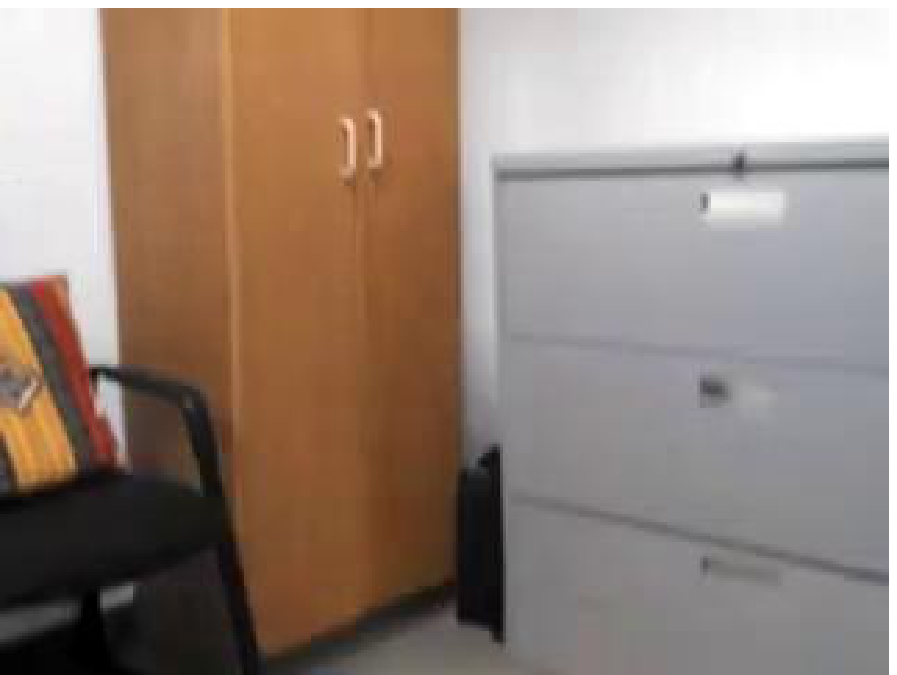}&
\includegraphics[width=0.1\textwidth]{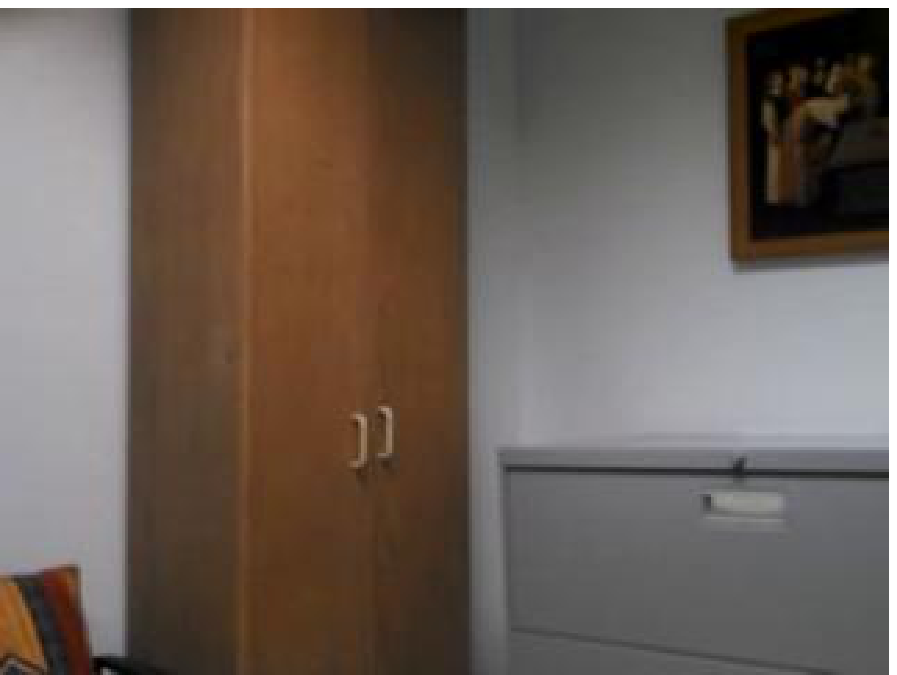}&
\includegraphics[width=0.1\textwidth]{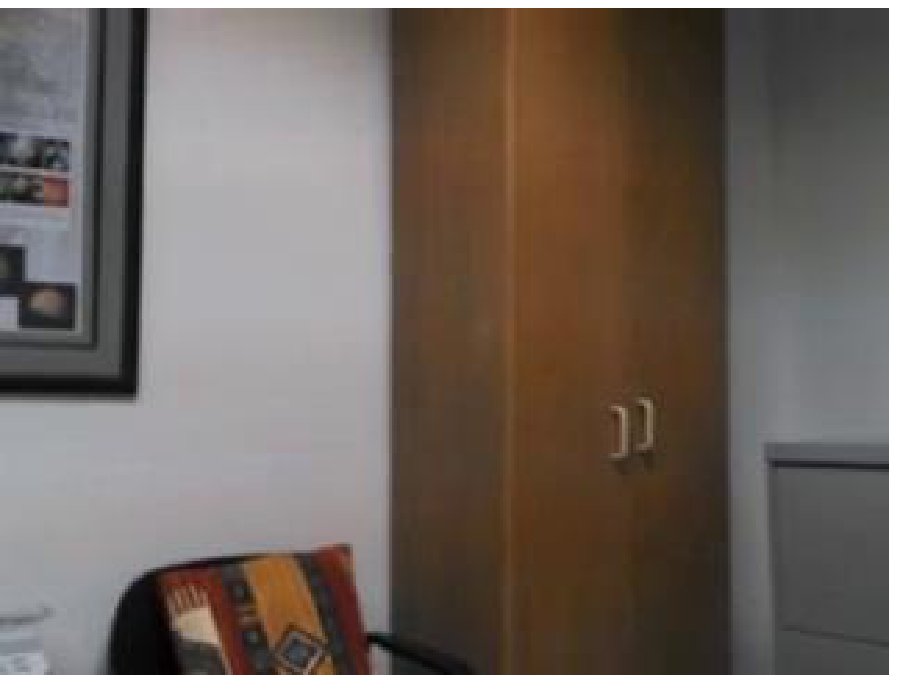}&
\includegraphics[width=0.1\textwidth]{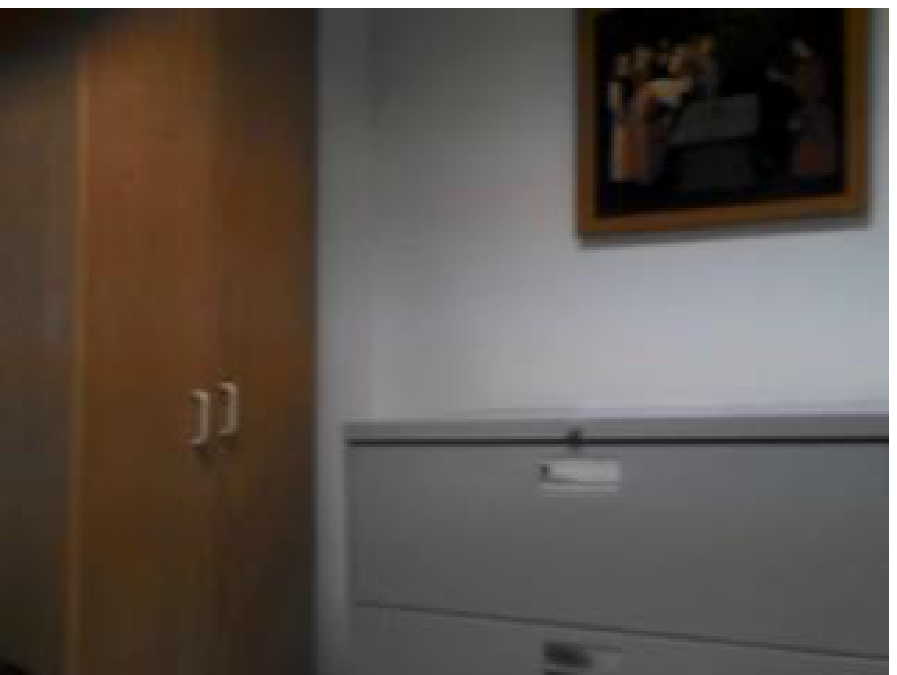}&
\includegraphics[width=0.1\textwidth]{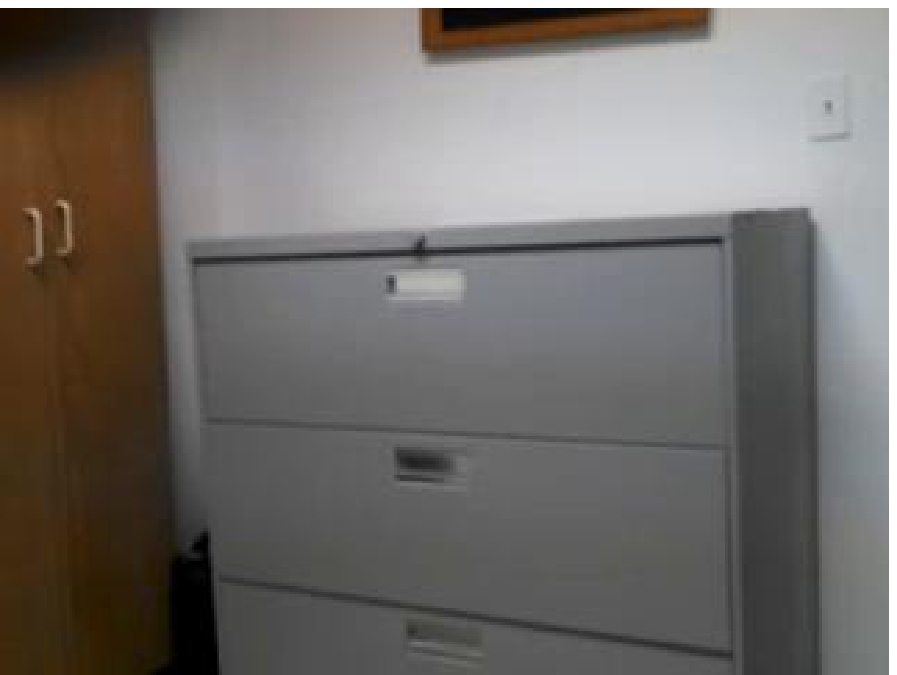}&
\includegraphics[width=0.1\textwidth]{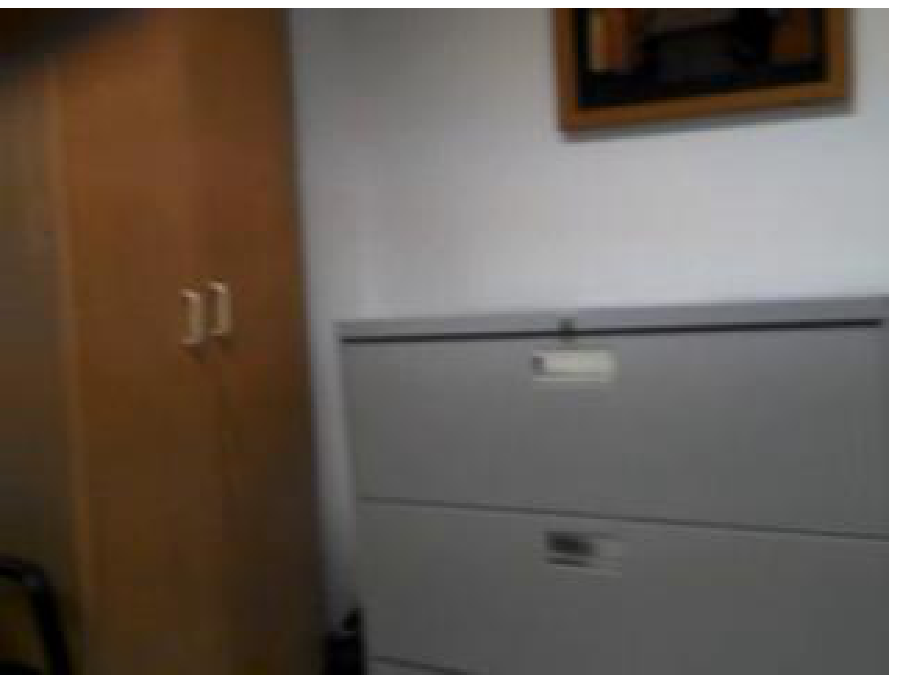}&
\includegraphics[width=0.1\textwidth]{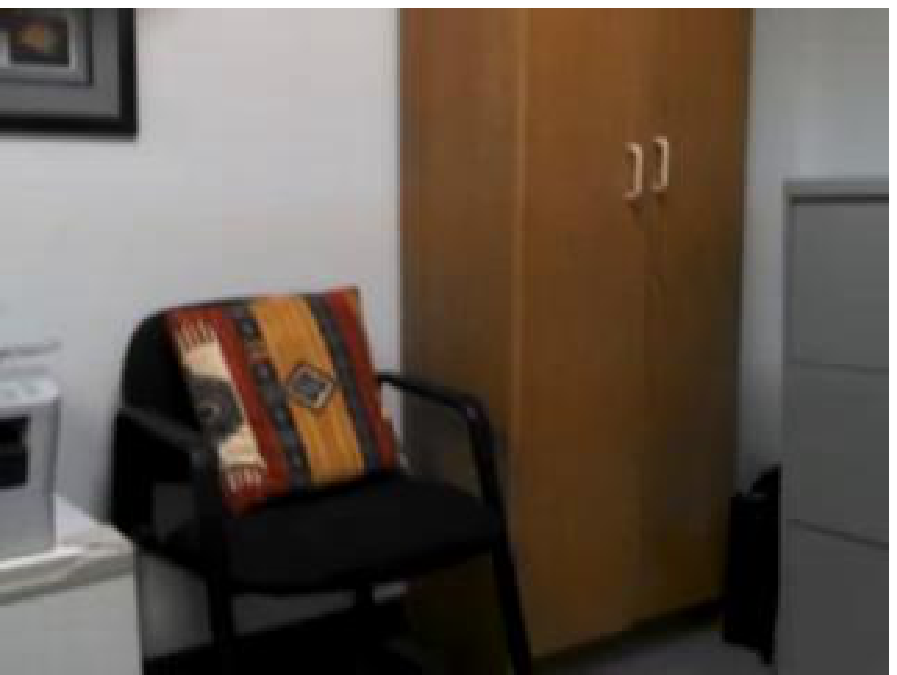}&
\includegraphics[width=0.1\textwidth]{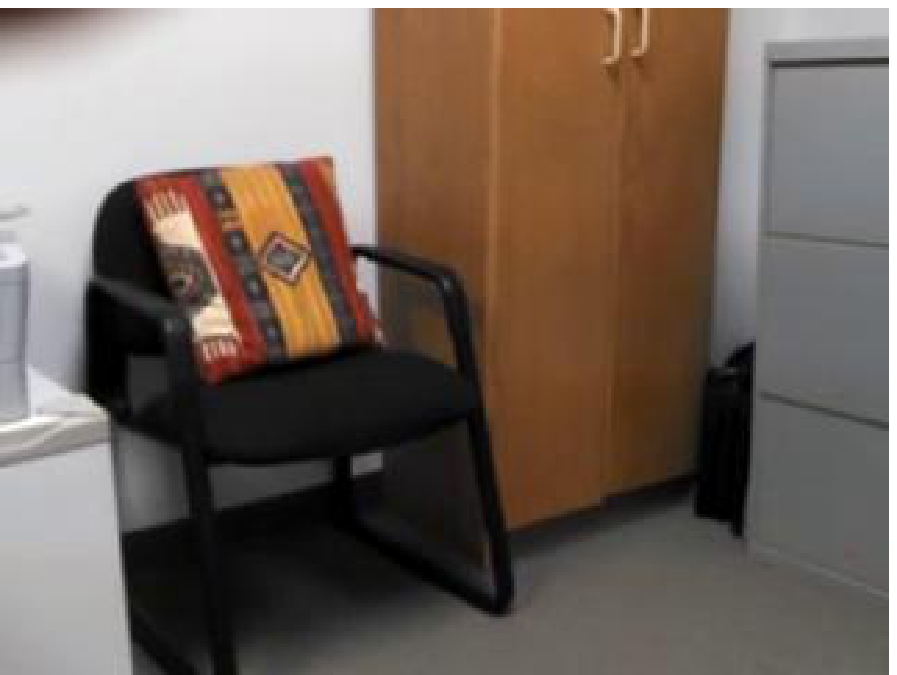}&
\includegraphics[width=0.1\textwidth]{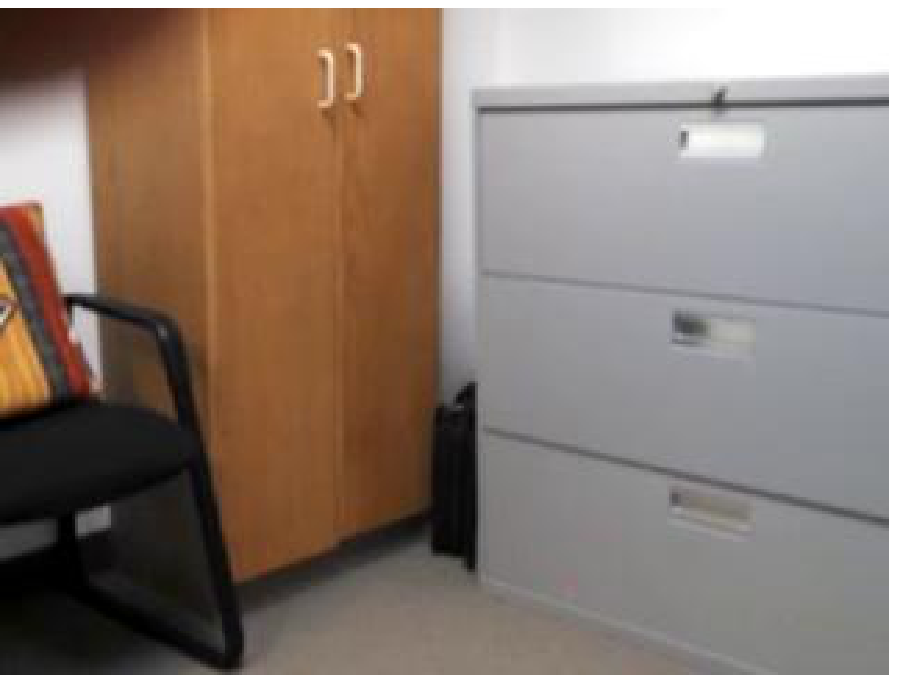}&
\includegraphics[width=0.1\textwidth]{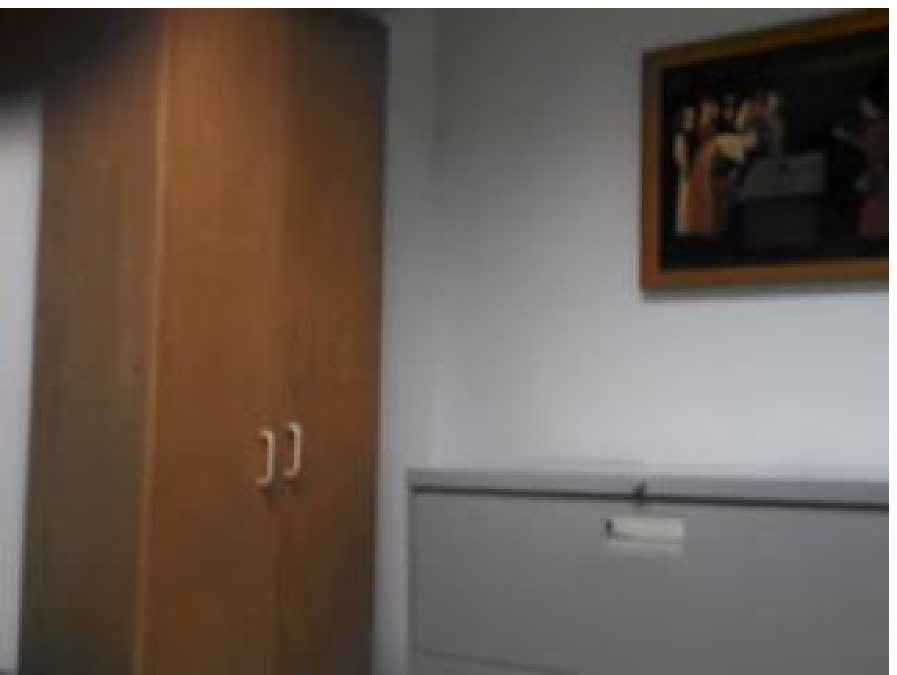}
\\
\includegraphics[width=0.1\textwidth]{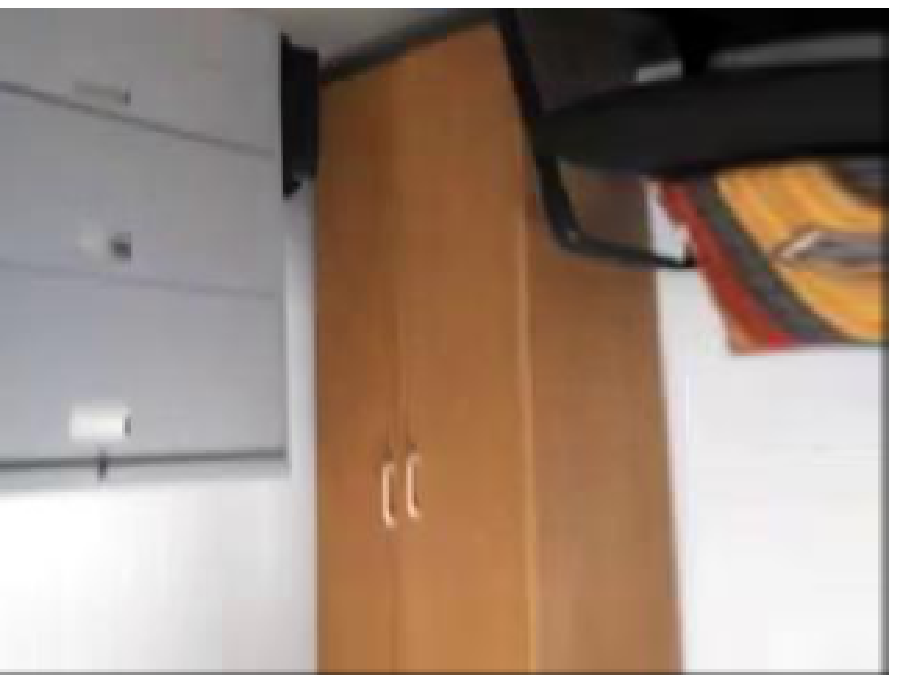}&
\includegraphics[width=0.1\textwidth]{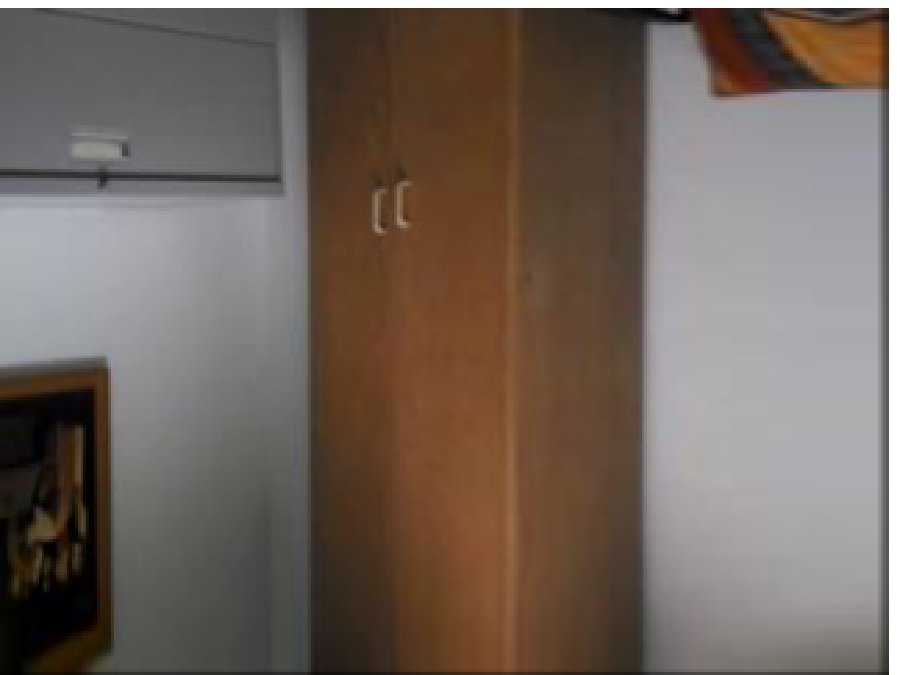}&
\includegraphics[width=0.1\textwidth]{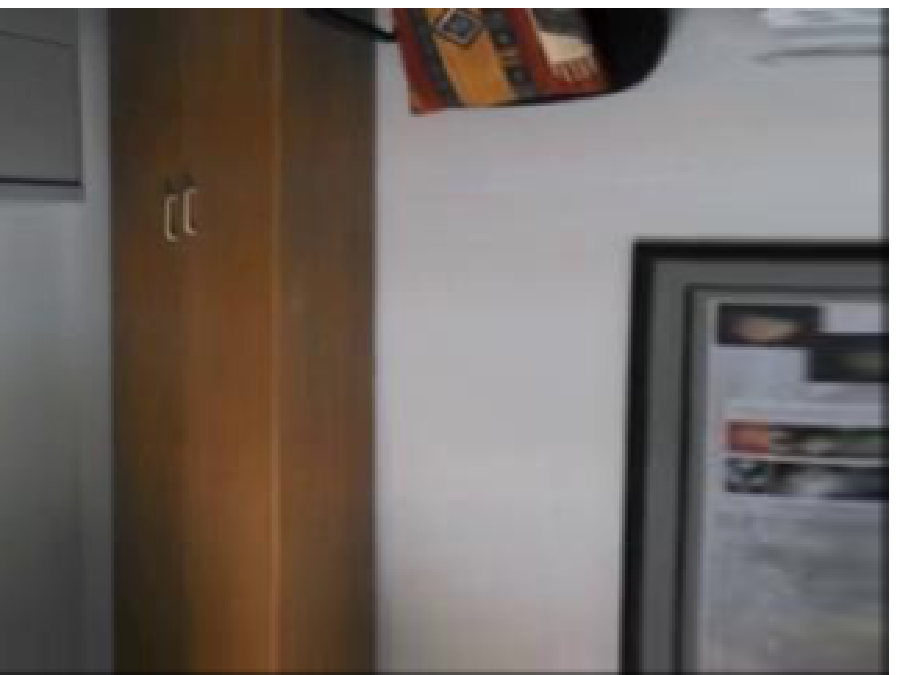}&
\includegraphics[width=0.1\textwidth]{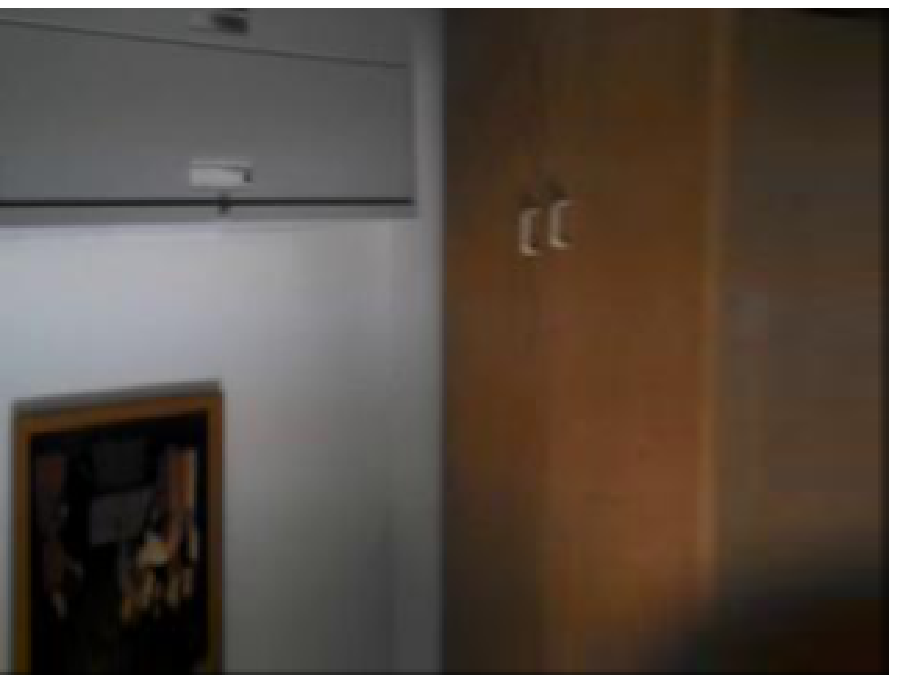}&
\includegraphics[width=0.1\textwidth]{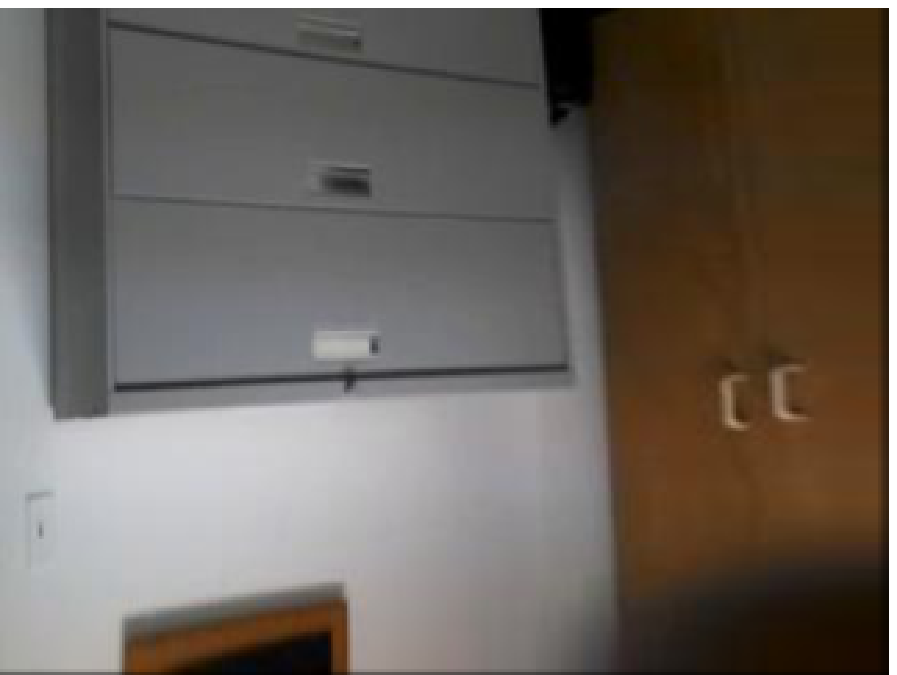}&
\includegraphics[width=0.1\textwidth]{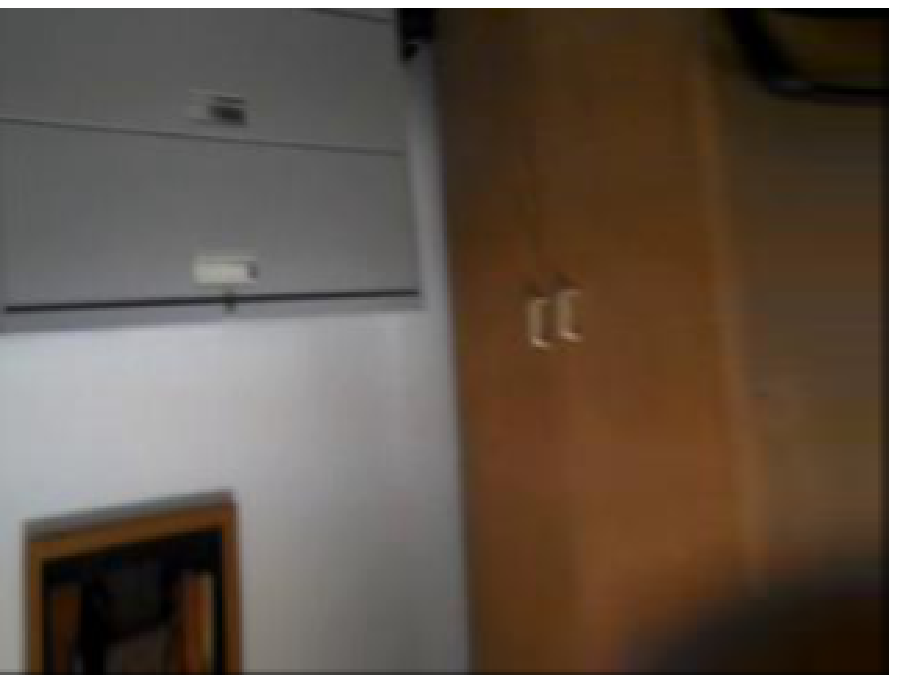}&
\includegraphics[width=0.1\textwidth]{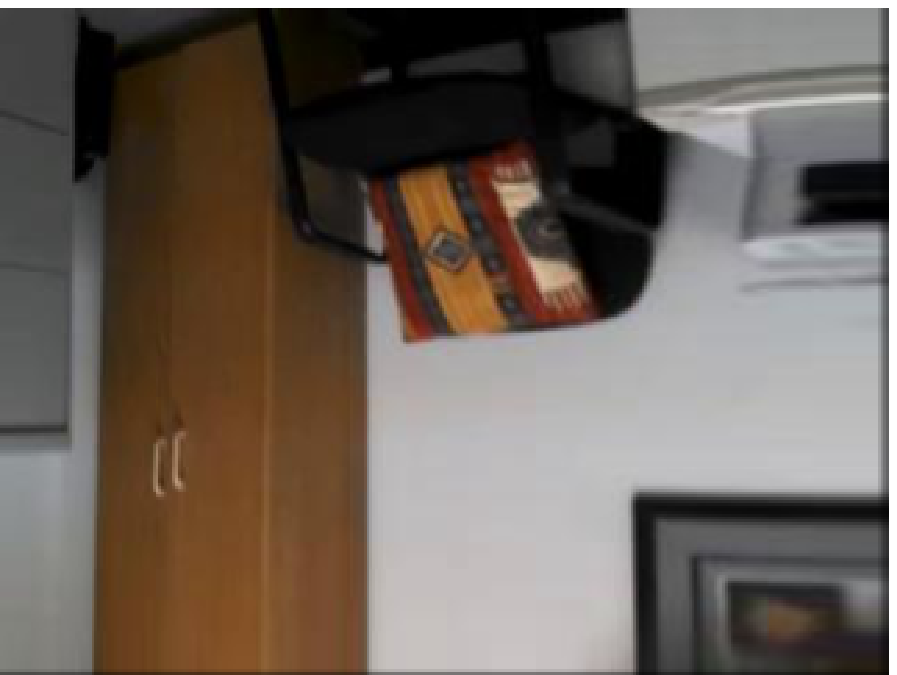}&
\includegraphics[width=0.1\textwidth]{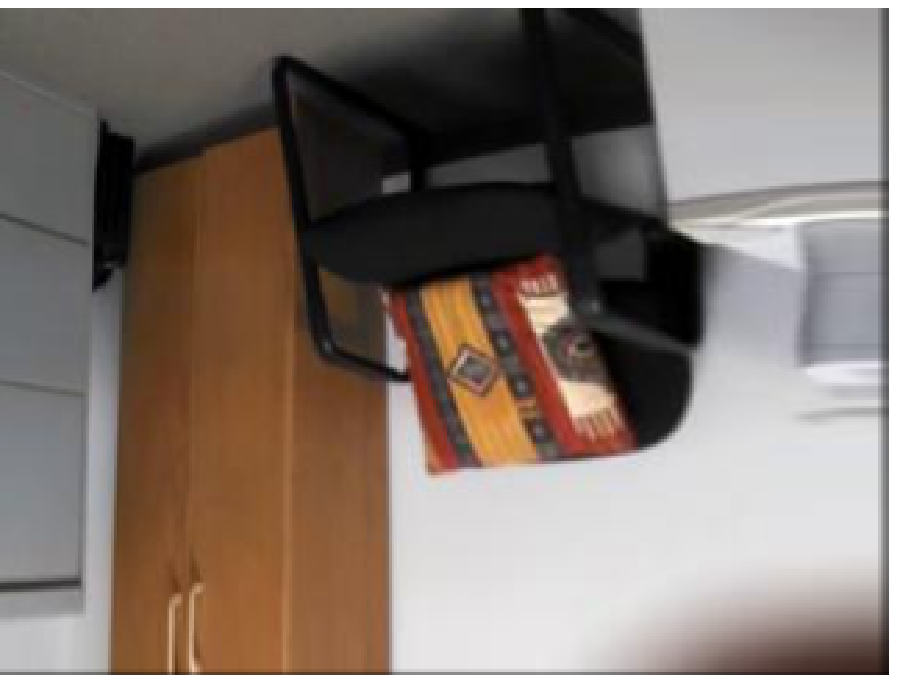}&
\includegraphics[width=0.1\textwidth]{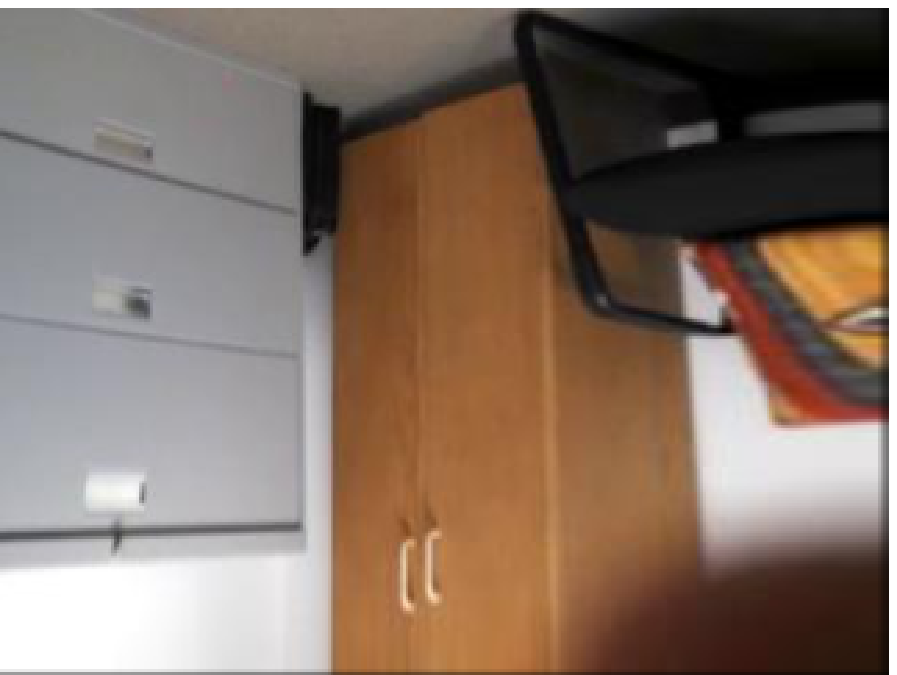}&
\includegraphics[width=0.1\textwidth]{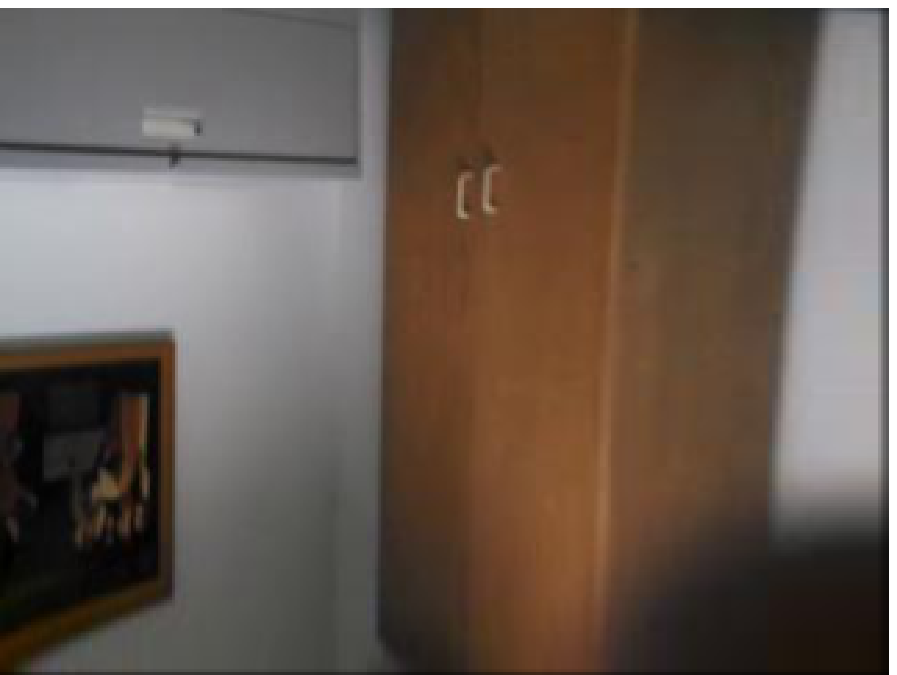}
\end{array}$
\end{center}
\caption{The top row shows sample images, which were recorded by a moving video camera at 30 frames per second and then subsampled at 1.67 s intervals. The bottom row show the images in the top row, after they were subjected to the nonlinear transformation depicted in Figure \ref{fig_hVOfHPrimeVPrime}}
\label{fig_frames}
\end{figure}

The original (i.e., untransformed) images were recorded by a cell phone video camera as it was moved in an irregular fashion over a portion of a spherical surface, having a radius of approximately 25 cm. The plane of the camera was oriented so that it was always tangential to the surface, and the camera's lower edge was kept parallel to the floor at all times. In this way, the camera was moved with two degrees of freedom; i.e., it was moved through a series of configurations (positions and orientations) that formed a two-dimensional manifold. The camera recorded thirty frames per second over the course of approximately 70 minutes, producing a total of 126,036 frames. Each frame consisted of a $320 \times 240$ array of pixels, in which the RGB responses were measured with one byte of depth. The top row of Figure \ref{fig_frames} displays a typical series of images, subsampled at 1.67 s intervals over the course of 17 s.

\begin{figure}
\centering
\subfloat[]{\includegraphics[width=0.35\textwidth]{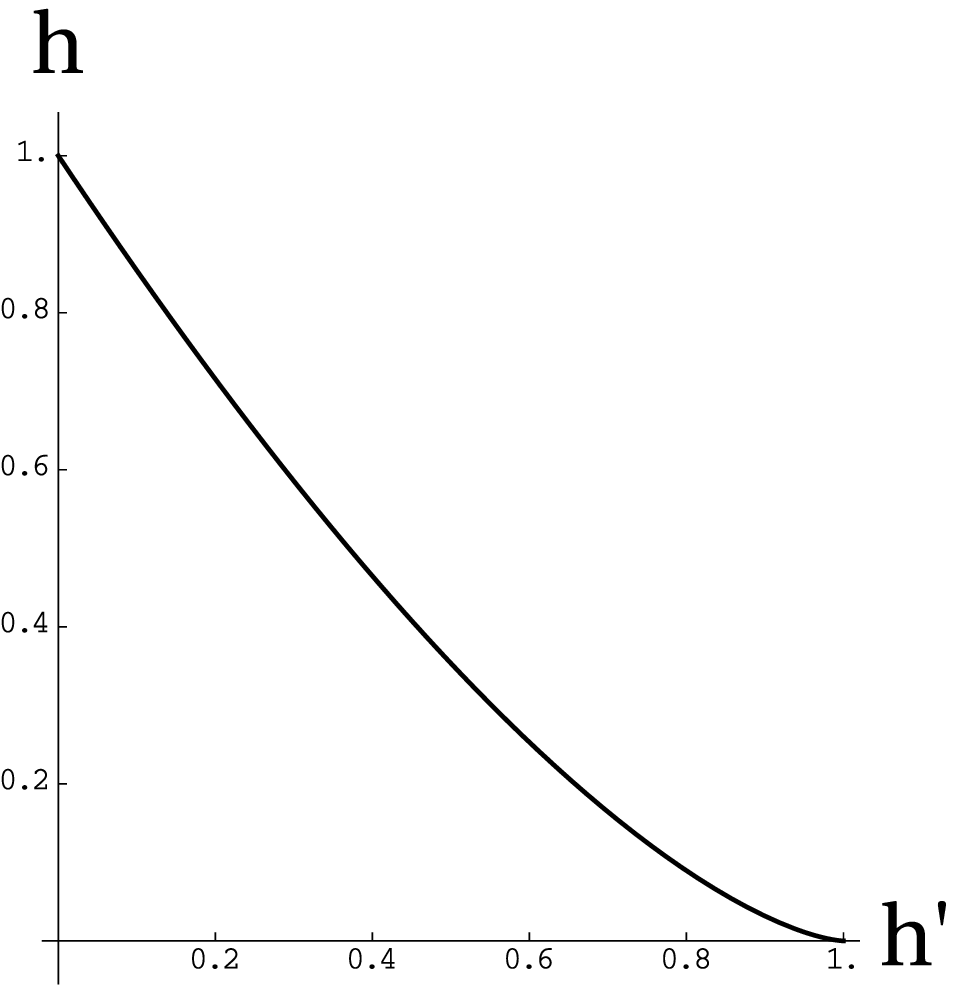}%
}
\subfloat[]{\includegraphics[width=0.35\textwidth]{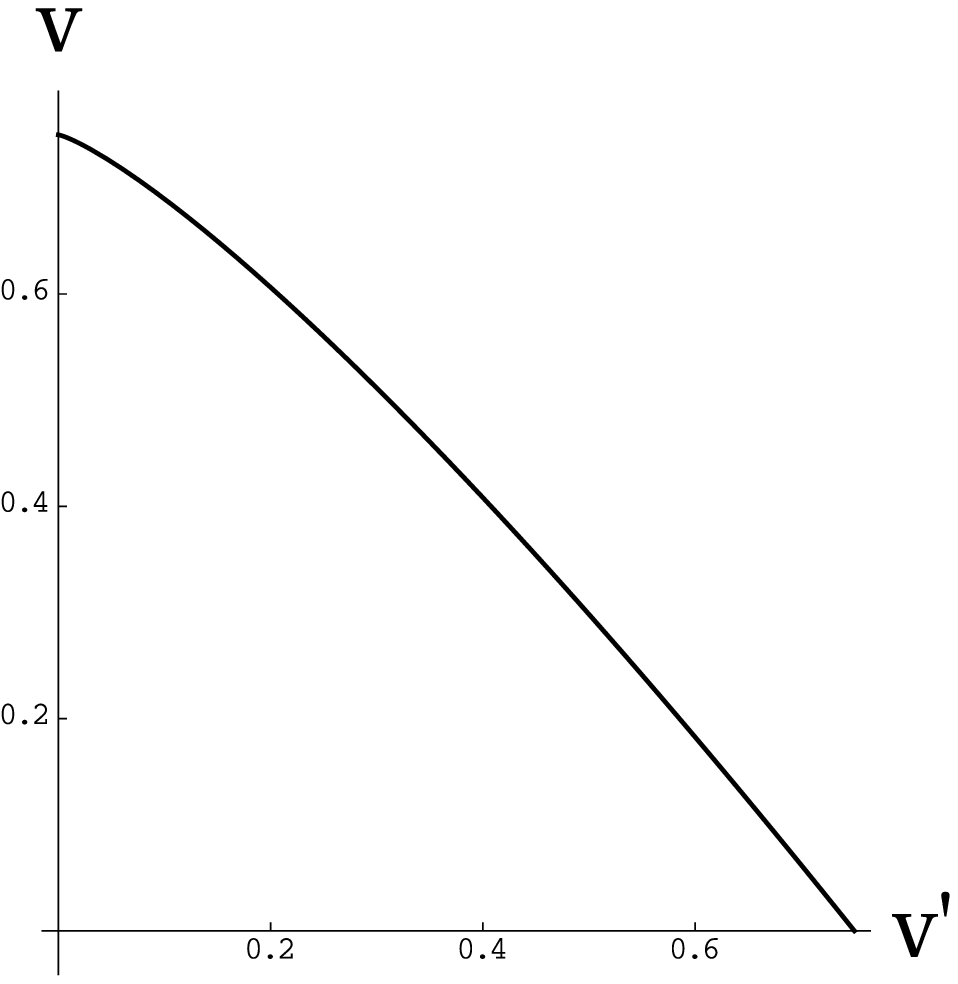}%
}
\caption{The nonlinear transformation between $(h,v)$, the coordinates of a pixel in each recorded image, and $(h',v')$, the coordinates of the corresponding pixel in the transformed image.}
\label{fig_hVOfHPrimeVPrime}
\end{figure}

The second time series of images was created by subjecting each recorded image to a nonlinear transformation.  Specifically, each pixel with image coordinates $(h,v)$ in a given recorded frame was mapped to the location with image coordinates $(h'(h),v'(v))$ in the corresponding transformed frame, where $h(h')$ and $v(v')$ are shown in Figure \ref{fig_hVOfHPrimeVPrime}. It is evident that this transformation turns each image upside down and backwards, in addition to stretching or compressing each image near its borders. The bottom row of Figure \ref{fig_frames} shows the images that were produced by nonlinearly transforming the corresponding recorded frames in the top row. These images simulate the output of a different sensor (e.g., a video camera, which was "wearing" goggles having inverting/distorting lenses).

\begin{figure}
\centering
\subfloat[]{\includegraphics[width=0.35\textwidth]{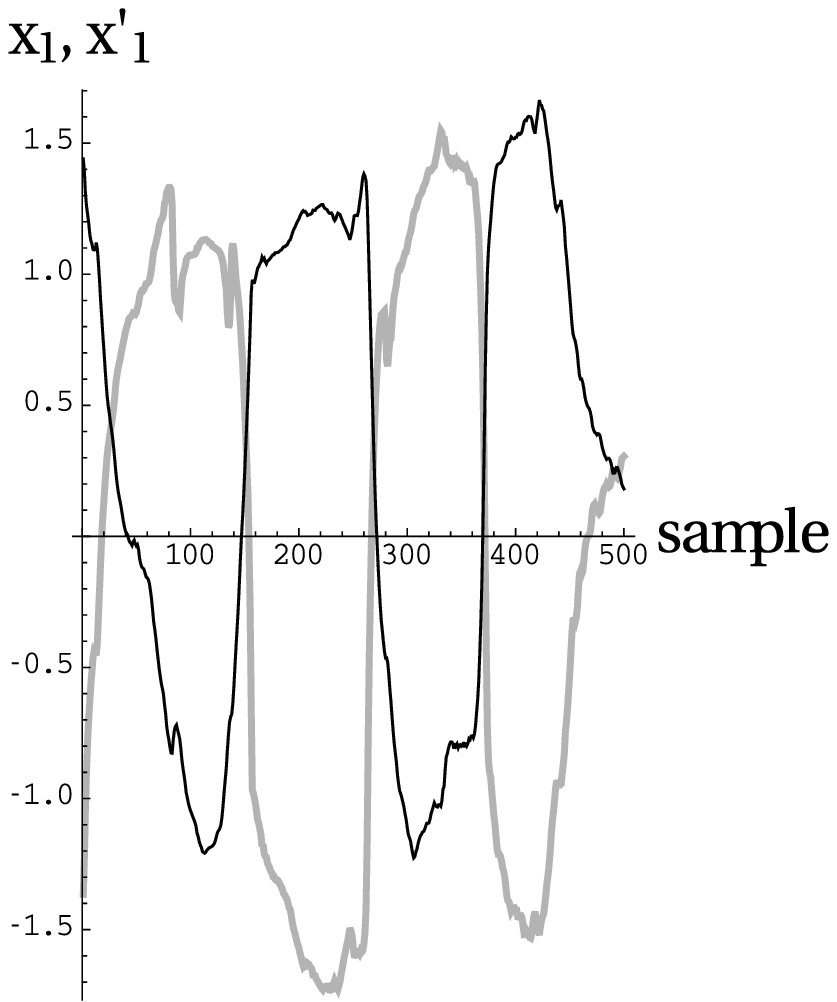}%
}
\subfloat[]{\includegraphics[width=0.35\textwidth]{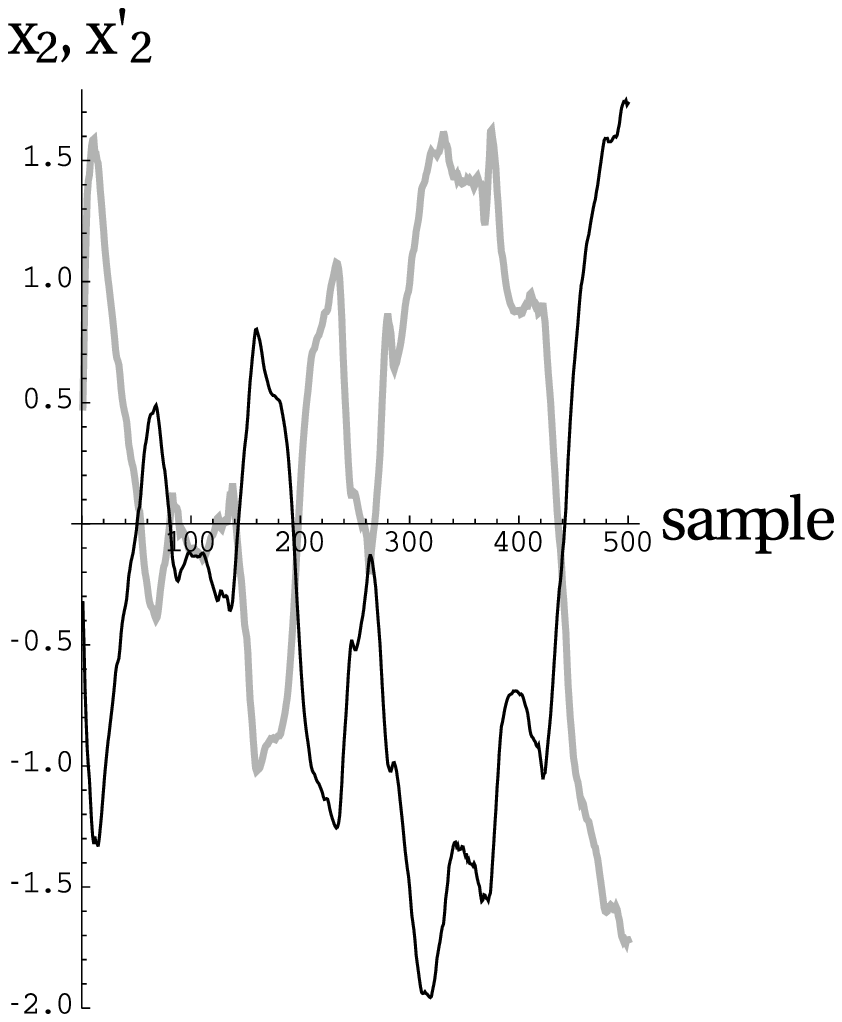}%
}
\caption{The thin black lines and the thick gray lines show the sensor measurements, $x(t)$ and $x'(t)$, derived from the sequences of untransformed and transformed images, respectively, during a 17 s time interval.}
\label{fig_xXPrimeOfT_video}
\end{figure}

Because the video was recorded as the camera moved through a two-dimensional manifold of configurations, the resulting images were expected to form a two-dimensional manifold in which each frame was represented by a point. A coordinate system, $x$, was imposed on this manifold in the following manner. First, we computed six numbers consisting of the centroids of the R, G, and B components for each recorded image. Then, we did a principal components analysis of the collection of six-dimensional multiplets for all recorded video frames. This showed that these multiplets were in or close to a two-dimensional planar subspace, which contained $99\%$ of their variance. Because this subspace did not self-intersect, its points were invertibly related to the configurations of the camera. The $x$ coordinates of each image were taken to be the first two variance-normalized principal components of the corresponding multiplet. The same procedure was applied to the collection of transformed images in order to assign a two-component coordinate, $x'$, to each one. The thin black lines in Figure \ref{fig_xXPrimeOfT_video} show the measurement time series, $x(t)$, derived from the images recorded during a typical 17 s time interval. The thick gray lines in the same figure show the sensor measurements, $x'(t)$, derived from the sequence of transformed images during the same time interval. The $x(t)$ and $x'(t)$ time series can be considered to be the measurements that were produced by two observers who were watching the same physical system with different sensors (i.e., with an ordinary video camera and with a camera having distorting/inverting lenses, respectively). Alternatively, $x'(t)$ can be considered to be the measurements $x(t)$, after they have been transformed to another coordinate system ($x'$) on the two-dimensional manifold of images.

\begin{figure}
\centering
\subfloat[]{\includegraphics[width=0.35\textwidth]{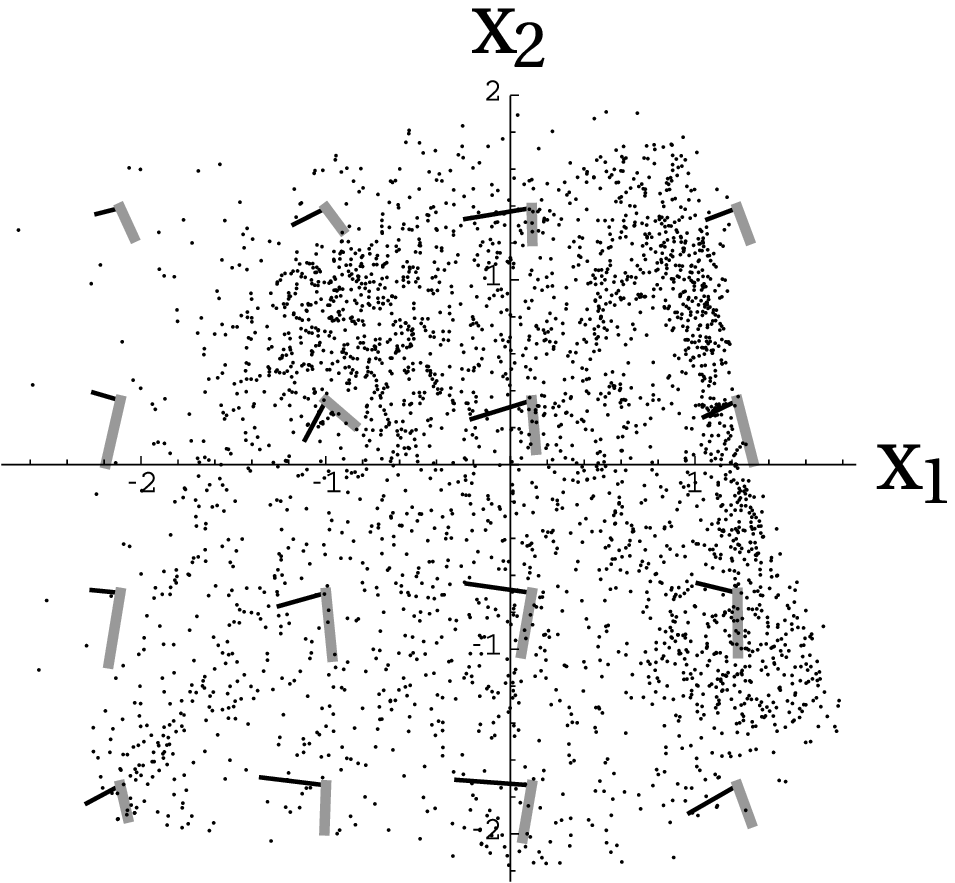}%
}
\subfloat[]{\includegraphics[width=0.35\textwidth]{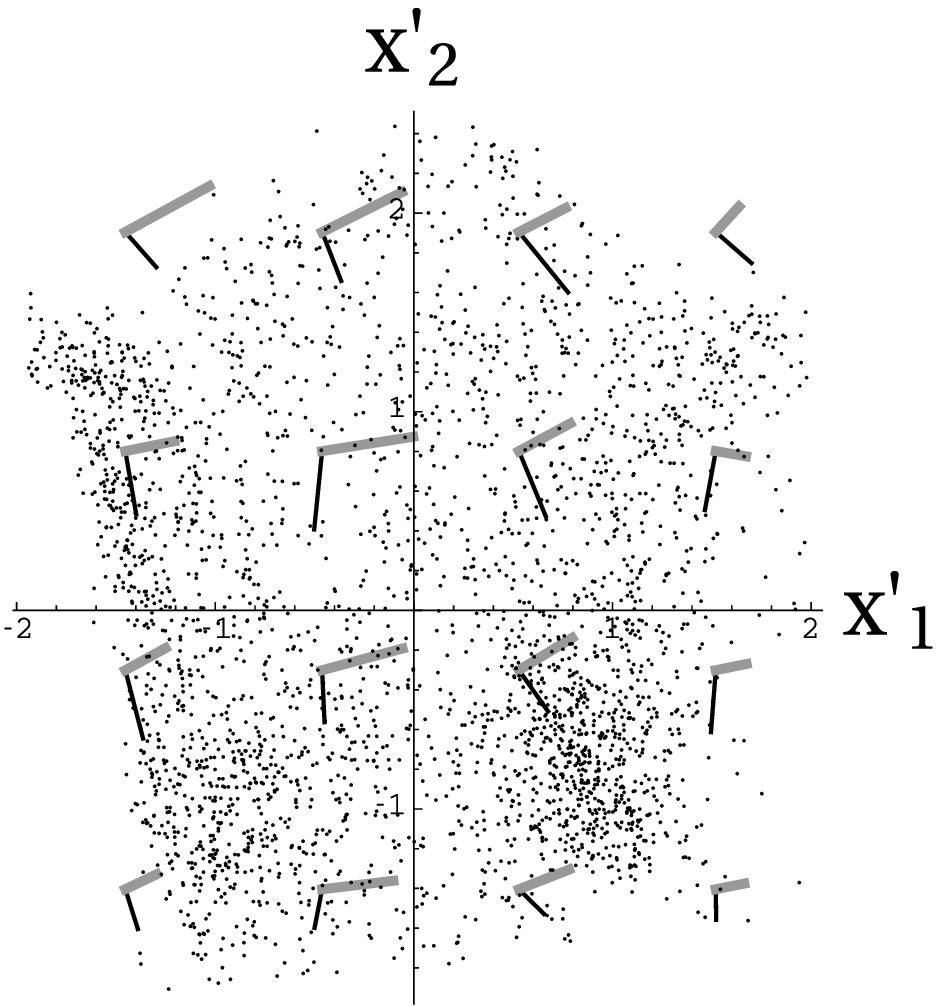}%
}
\caption{The thin black lines and the thick gray lines in the left panel show the local vectors, $V_{(1)}$ and $V_{(2)}$, respectively, derived from the sequence of untransformed images. The right panel shows the local vectors derived from the transformed images. The black points in each panel show the coordinates of a random sample of the measurements, $x(t)$ and $x'(t)$, derived from the recorded and transformed images, respectively.}
\label{fig_VOfXVPrimeOfXPrime_video}
\end{figure}

The 126,036 measurements, $x(t)$, derived from the sequence of untransformed images, were assigned to bins in a $4 \times 4$ array. Then, the procedure in Subsection \ref{derivation} was used to compute the local vectors in each bin ($V_{(i)}(x)$ for $i=1,2$). The same procedure was applied to measurements $x'(t)$, derived from the transformed images, in order to compute the local vectors, $V'_{(i)}(x')$. These local vectors are shown in the left and right panels of Figure \ref{fig_VOfXVPrimeOfXPrime_video}.

\begin{figure}
\centering
\subfloat[]{\includegraphics[width=0.35\textwidth]{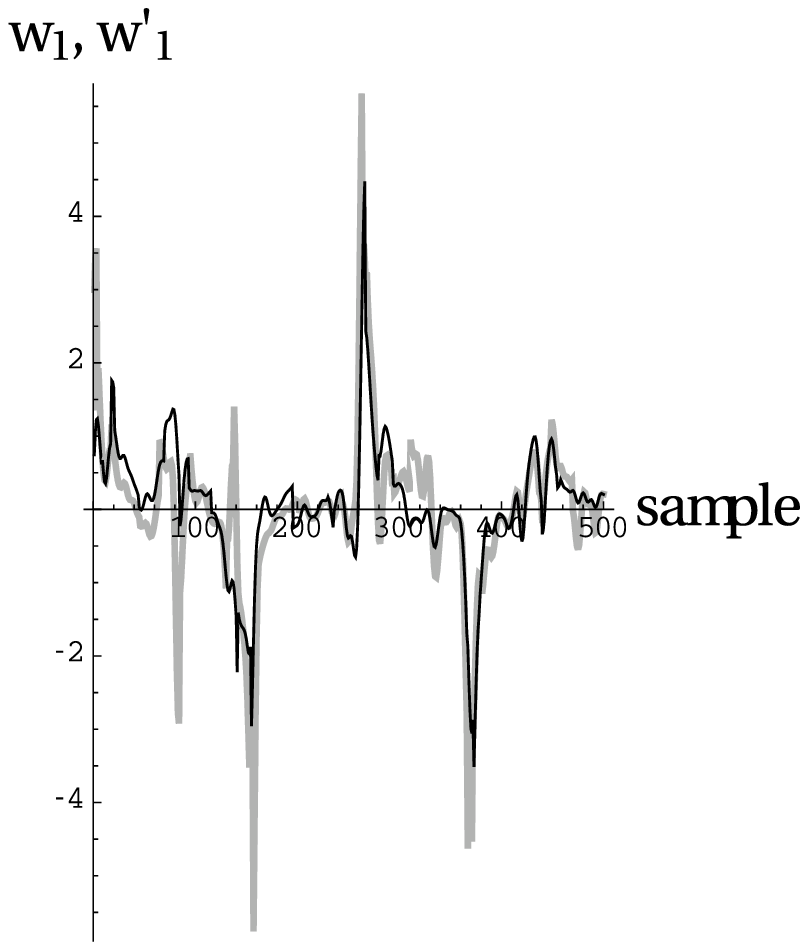}%
}
\subfloat[]{\includegraphics[width=0.35\textwidth]{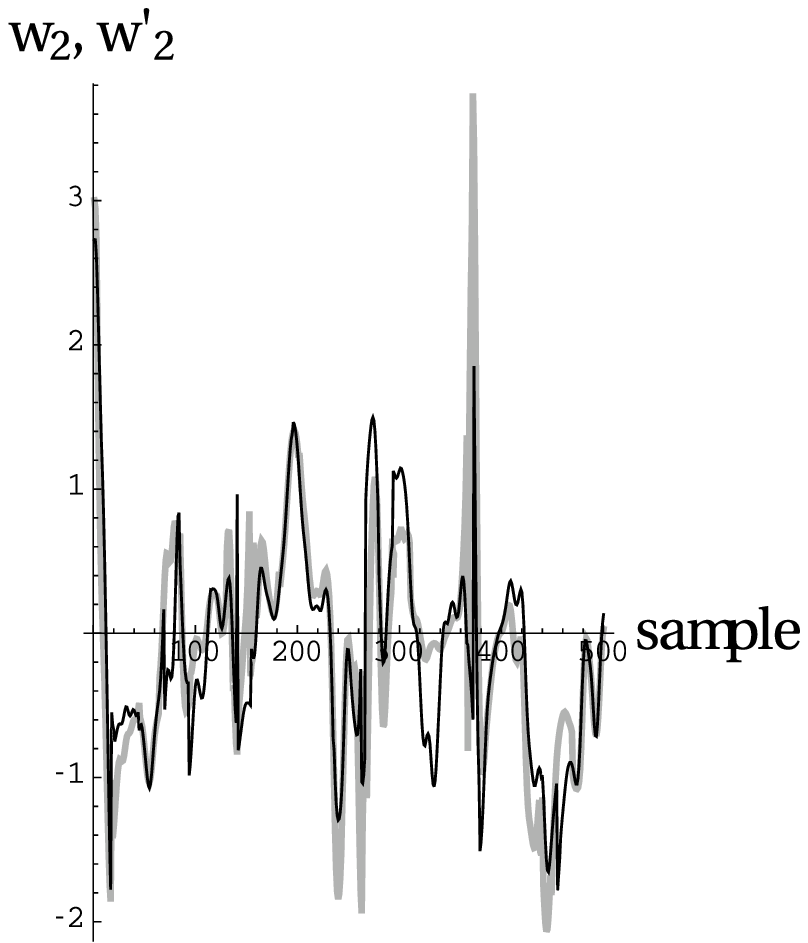}%
}
\caption{The thin black lines and the thick gray lines show the inner time series, $w_{i}(t)$ and $w'_{i}(t)$, derived from the sequences of untransformed and transformed images, respectively, during the 17 s time interval depicted in Figure \ref{fig_xXPrimeOfT_video}.}
\label{fig_wWPrimeOfT_video}
\end{figure}

The measurement time series, $x(t)$, and the corresponding local vectors, $V_{(i)}$, were substituted in (\ref{xDot rep}) in order to derive the inner time series, $w_{i}(t)$, corresponding to the sequence of untransformed images. Likewise, the measurement time series, $x'(t)$, and the corresponding local vectors, $V'_{(i)}$, were used to derive the inner time series, $w'_{i}(t)$, corresponding to the sequence of transformed images. The thin black lines and the thick gray lines in Figure \ref{fig_wWPrimeOfT_video} show the weights, $w_{i}(t)$ and $w'_{i}(t)$, respectively, during the time interval depicted in Figure \ref{fig_xXPrimeOfT_video}, after $w'_{i}(t)$ was multiplied by a global permutation and reflection. Notice that the inner time series are nearly the same, despite the fact that they were derived from the outputs of dramatically different sensors. In other words, the inner time series are sensor-independent, as proved in Subsection \ref{derivation}. 

These results loosely mimic the findings of the well-known psychophysical experiments (\cite{Held}) in which subjects, who wore inverting/distorting goggles, eventually learned to perceive the world as it was perceived before wearing the goggles. Similarly, Figure \ref{fig_wWPrimeOfT_video} shows that the observer, whose camera was ''wearing" goggles, perceived the inner properties of the image time series to be the same (thick gray lines) as they were perceived before wearing the goggles (thin black lines).

\subsection{Nonlinear mixtures of two audio waveforms}
\label{two audio signals}

In this subsection, the system consists of two speakers, whose utterances are statistically independent and are observed in two ways: 1) as a pair of unmixed signals, each one being one speaker's waveform; 2) as a pair of nonlinear mixtures of the unmixed signals. The unmixed and mixed pairs of signals simulate measurements made by two observers who were using different sensors. The procedure in Subsection \ref{derivation} was applied to derive the inner time series, corresponding to the unmixed and mixed signals. These inner time series are shown to be almost the same, thereby demonstrating their sensor independence. Furthermore, the time series of each weight component, derived from the signal mixtures, is almost the same as the time series of a weight component, derived from one of the unmixed signals. This demonstrates that the inner time series of a composite system is simply a collection of the inner time series of its statistically independent subsystems, as proved in Subsection \ref{composite systems}.

\begin{figure}
\centering
\subfloat[]{\includegraphics[width=0.35\textwidth]{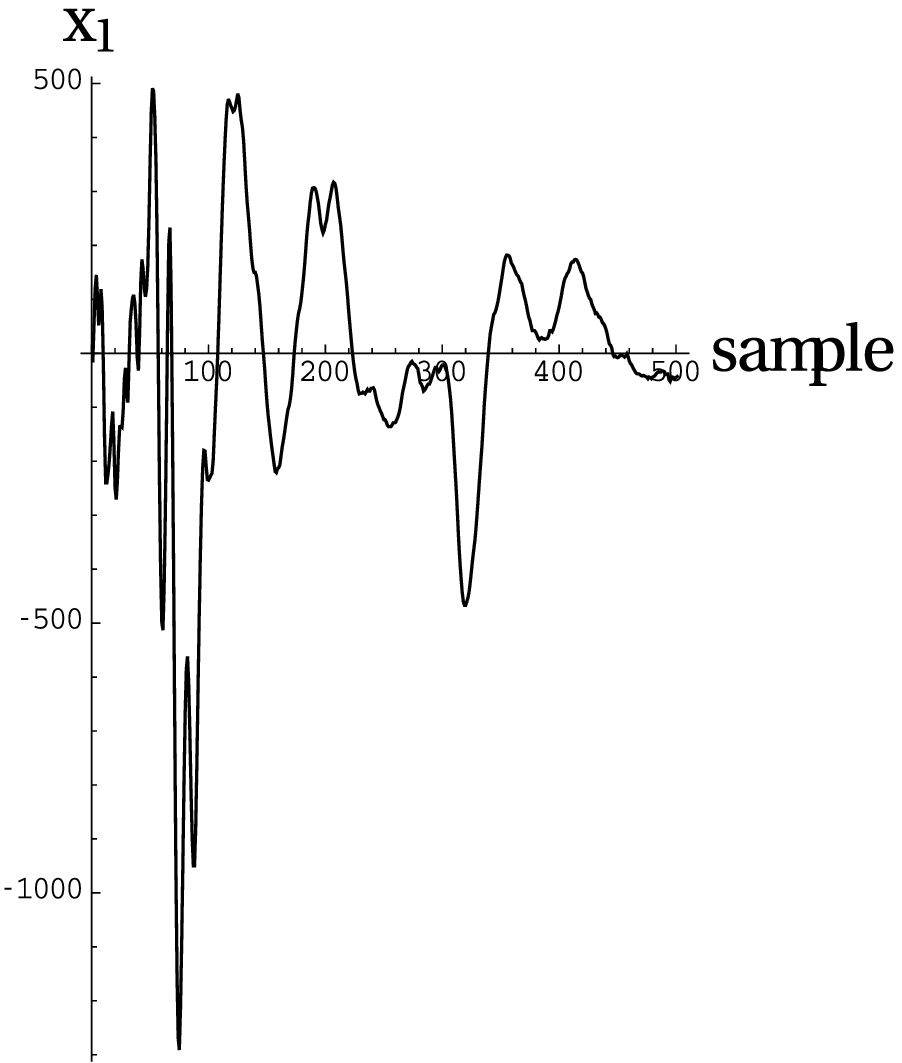}%
}
\subfloat[]{\includegraphics[width=0.35\textwidth]{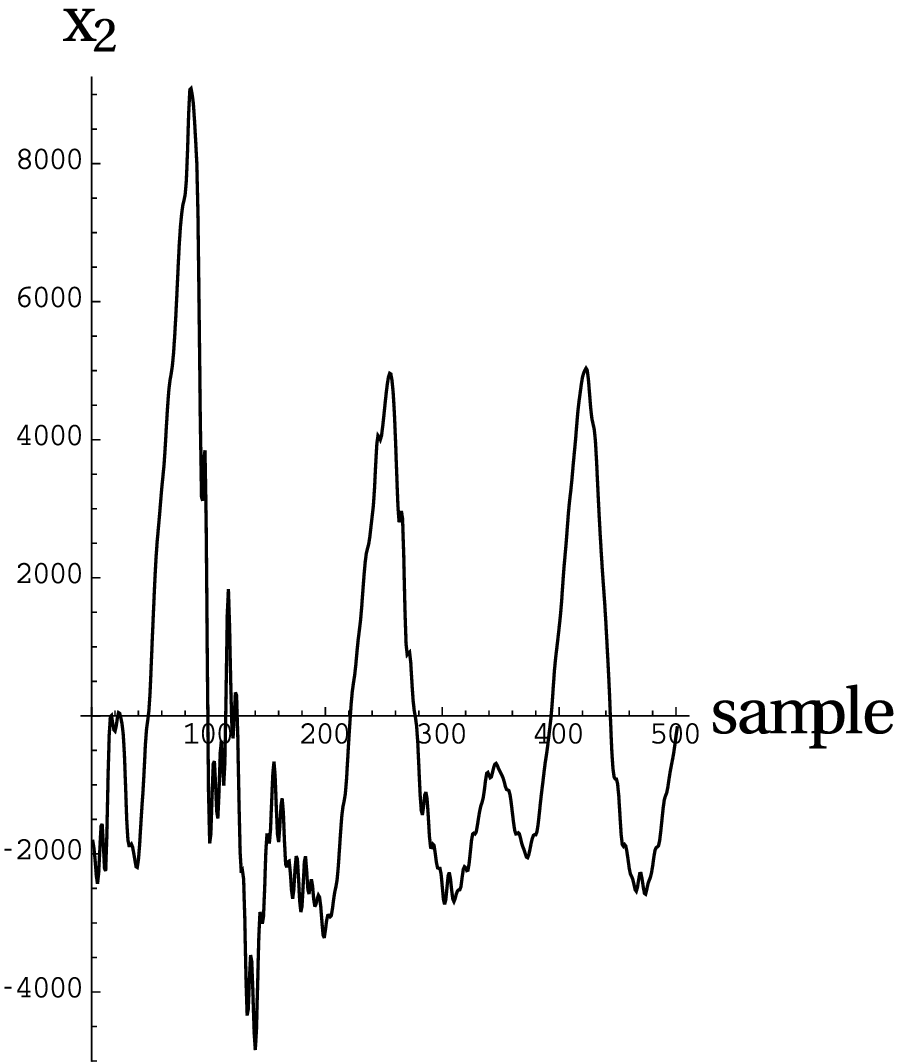}%
}
\caption{The unmixed audio waveforms of the two speakers during a 31.25 ms time interval.}
\label{fig_xOfT_audio}
\end{figure}

\begin{figure}
\centering
\subfloat{\includegraphics[width=0.35\textwidth]{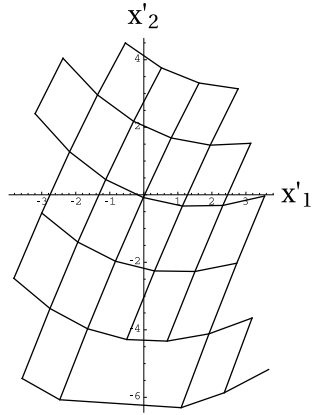}%
}
\caption{A warped grid in the $x'$ coordinate system, obtained by applying the nonlinear mixing function in (\ref{mixing}) to a regular Cartesian grid in the $x$ coordinate system.}
\label{fig_warpedGrid_audio}
\end{figure}

\begin{figure}
\centering
\subfloat[]{\includegraphics[width=0.35\textwidth]{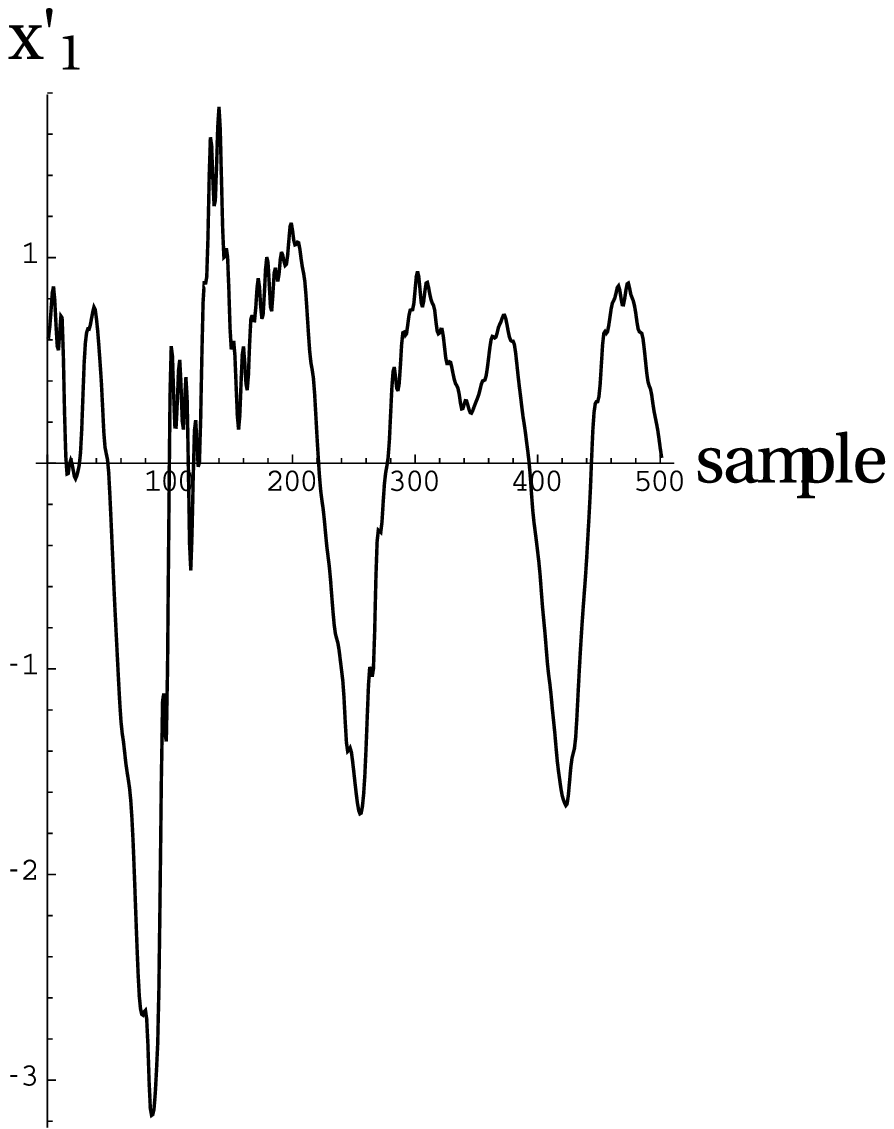}%
}
\subfloat[]{\includegraphics[width=0.35\textwidth]{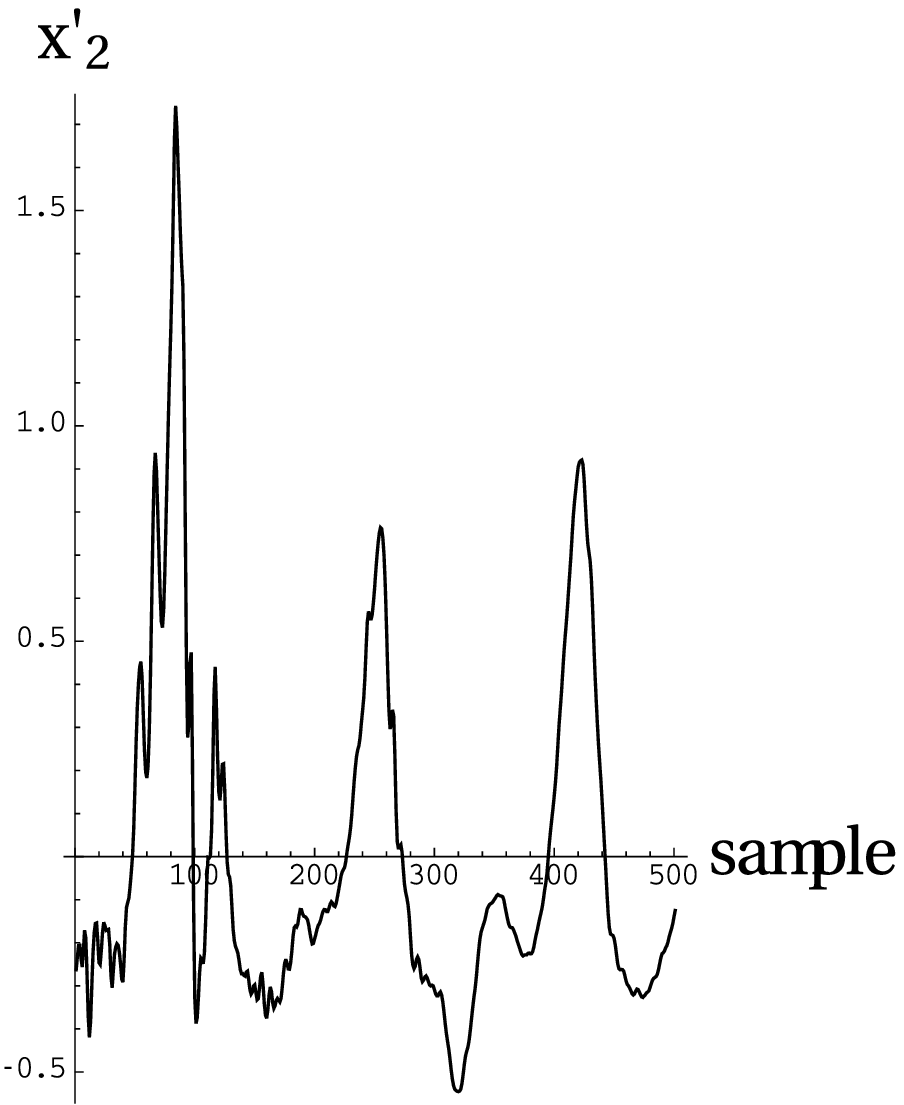}%
}
\caption{The mixed audio waveforms of the two speakers, obtained by applying the nonlinear mixing function in (\ref{mixing}) to the unmixed waveforms in Figure \ref{fig_xOfT_audio}.}
\label{fig_xPrimeOfT_audio}
\end{figure}

The unmixed signals were excerpts from audio book recordings of two male speakers, who were reading different texts. The two audio waveforms, denoted $x_{k}(t)$ for $k=1,2$, were 31.25 s long and were sampled 16,000 times per second with two bytes of depth. Figure \ref{fig_xOfT_audio} shows the two speakers' waveforms during a short (31.25 ms) interval. These waveforms were then  mixed by the nonlinear functions
\begin{equation}
\label{mixing}
\begin{split}
\mu_{1}(x) &= 0.763 x_1 + (958 - 0.0225 x_2)^{1.5} \\
\mu_{2}(x) &= 0.153 x_2 + (3.75 * 10^7-763 x_1 - 229 x_2)^{0.5} ,
\end{split}
\end{equation}
where $-2^{15} \leq s_1, s_2 \leq 2^{15}$. This is one of a variety of nonlinear transformations that were tried with similar results. The mixed measurements, $x'_{k}(t)$, were taken to be the variance-normalized, principal components of the waveform mixtures, $\mu_{k}[x(t)]$. Figure \ref{fig_warpedGrid_audio} shows how this nonlinear mixing function mapped an evenly-spaced Cartesian grid in the $x$ coordinate system onto a warped grid in the $x'$ coordinate system. Notice that the mapped grid does not ''fold over" onto itself, showing that it is an invertible mapping. The lines in Figure \ref{fig_xPrimeOfT_audio} show the time course of $x'(t)$. When either waveform mixture ($x'_{1}(t)$ or $x'_{2}(t)$) was played as an audio file, it sounded like a confusing superposition of two voices, which were quite difficult to understand. 

\begin{figure}
\centering
\subfloat[]{\includegraphics[width=0.25\textwidth]{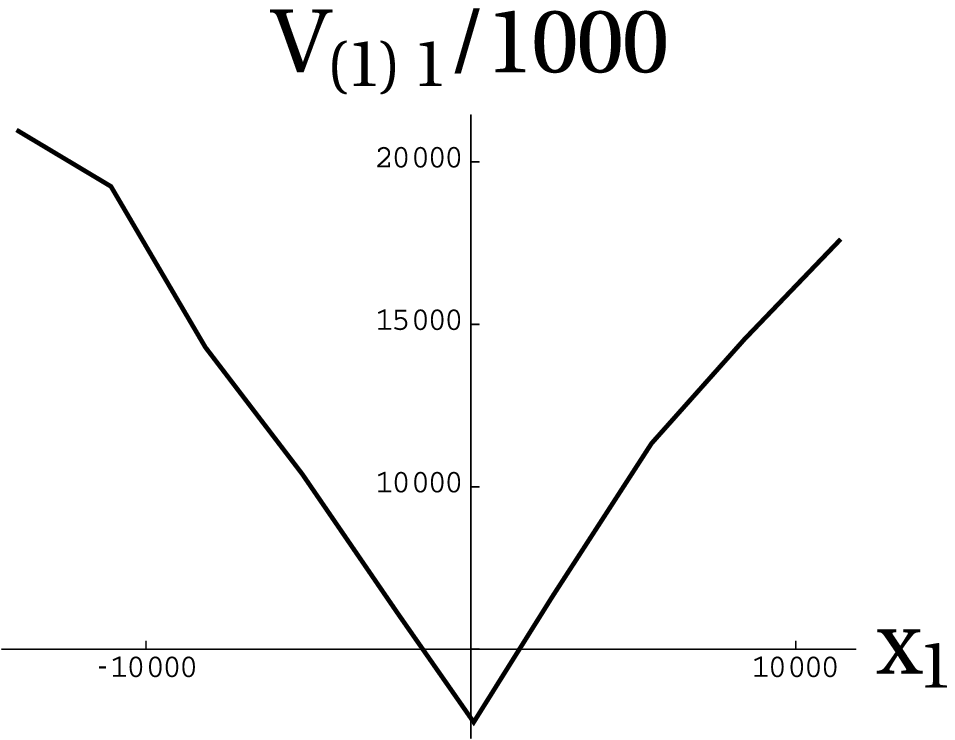}%
}
\subfloat[]{\includegraphics[width=0.25\textwidth]{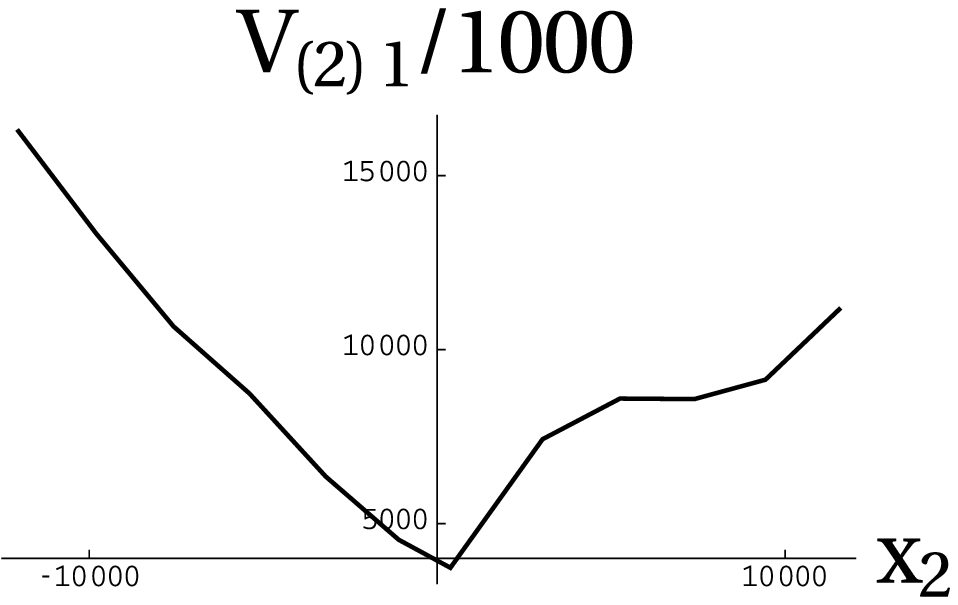}%
}
\subfloat[]{\includegraphics[width=0.25\textwidth]{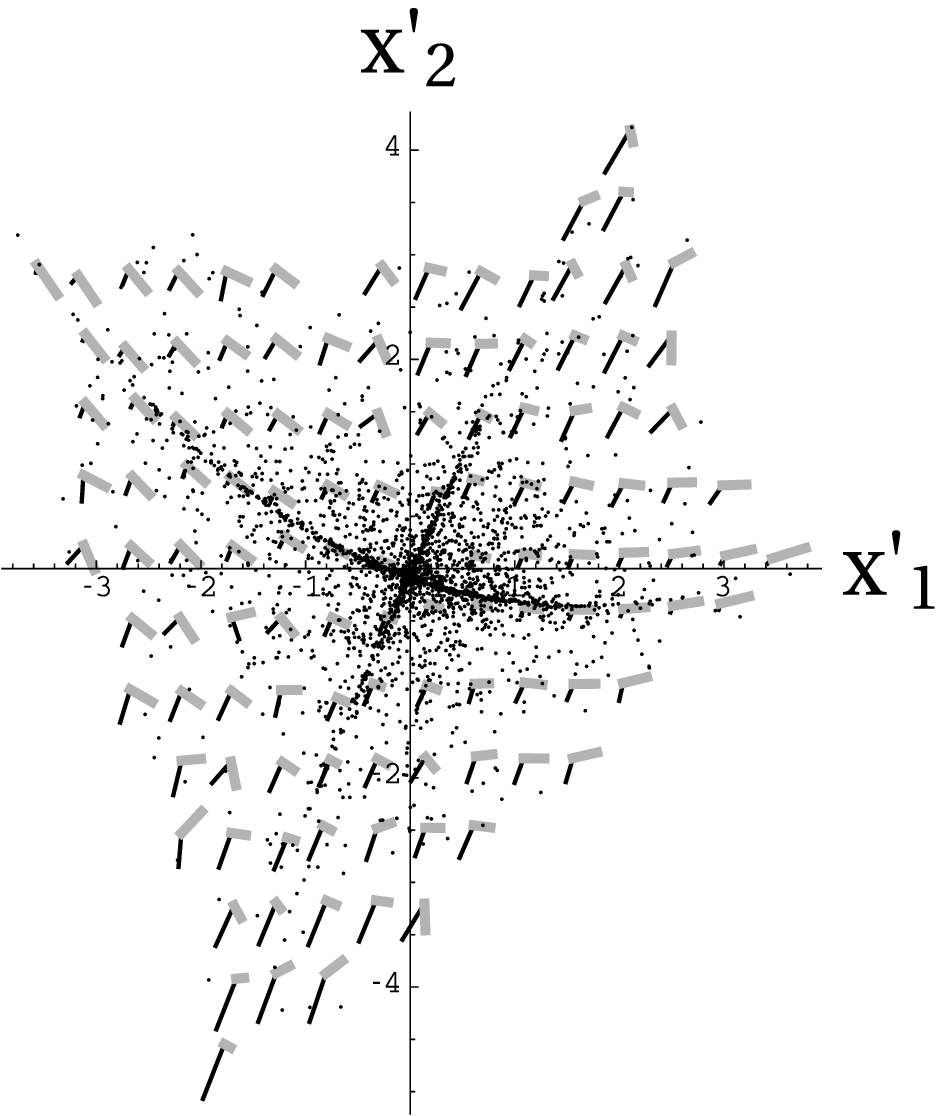}%
}
\caption{The left and middle panels show the one-component local vectors derived from the unmixed waveforms, $x_{1}(t)$ and $x_{2}(t)$, excerpts of which are illustrated in Figure \ref{fig_xOfT_audio}. The line segments in the right panel show the local vectors derived from the mixed waveforms, $x'(t)$, excerpts of which are illustrated in Figure \ref{fig_xPrimeOfT_audio}. These line segments have been uniformly rescaled for the purpose of display. The small black points in the right panel show the distribution of randomly chosen samples of the mixed waveforms, $x'(t)$.}
\label{fig_VVPrimeOFXXPrime_audio}
\end{figure}

\begin{figure}
\centering
\subfloat[]{\includegraphics[width=0.35\textwidth]{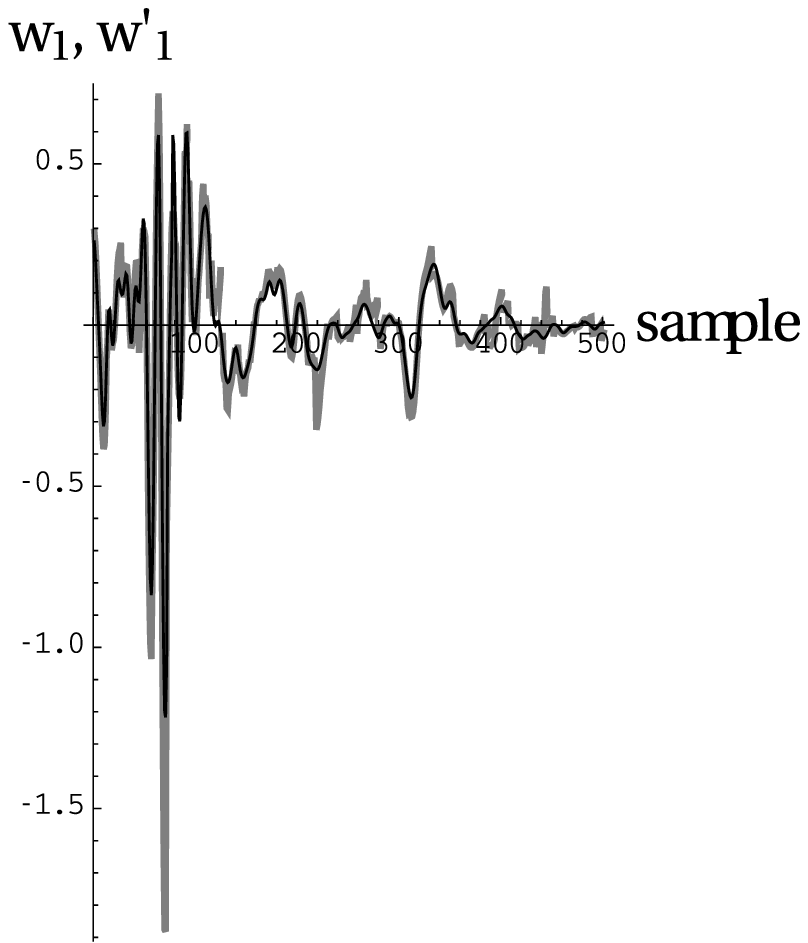}%
}
\subfloat[]{\includegraphics[width=0.35\textwidth]{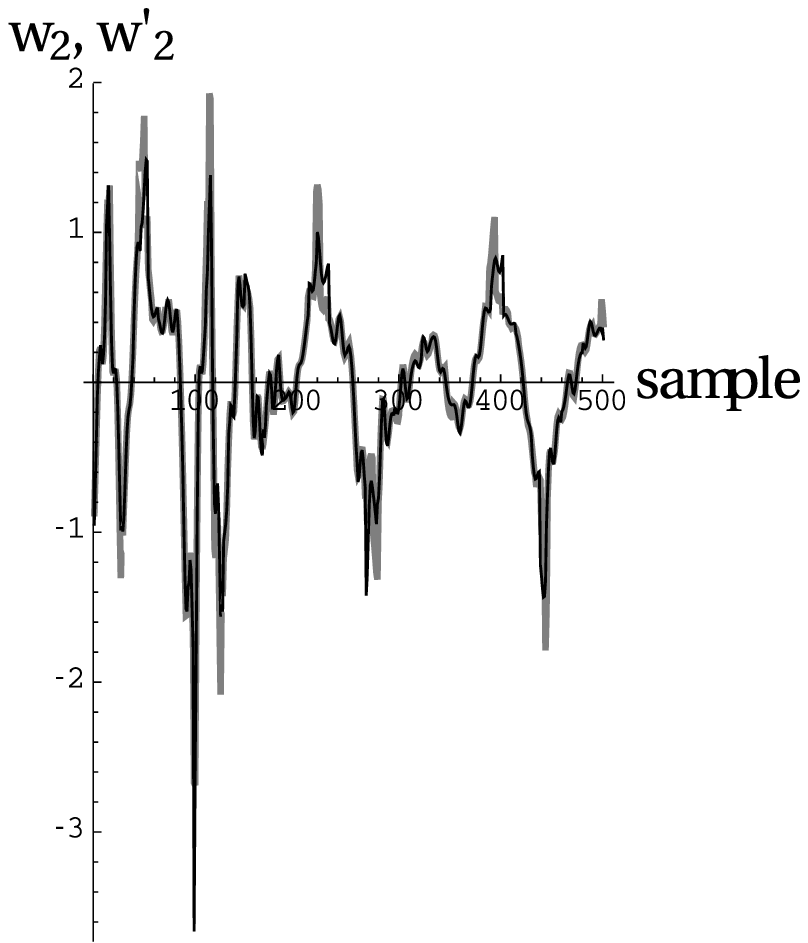}%
}
\caption{The thin black lines and the thick gray lines show the inner time series, $w_{i}(t)$ and $w'_{i}(t)$, derived from the unmixed and mixed waveforms, respectively, during the 31.25 ms time interval depicted in Figure \ref{fig_xOfT_audio} and \ref{fig_xPrimeOfT_audio}.}
\label{fig_wWPrimeOfT_audio}
\end{figure}

The method in Section \ref{method} was then applied to these data as follows:
\begin{enumerate}
\item The 500,000 measurements of the first unmixed waveform, consisting of $x_1$ and $\dot{x}_1$ at each sampled time, were sorted into an array of 16 bins in $x_1 \mbox{-space}$. Then, the $\dot{x}$ distribution in each bin was used to compute local velocity correlations, and these were used to derive the one-component local vector, $V_{(1)}(x_1)$, in each bin in $x_1 \mbox{-space}$. The left panel of figure \ref{fig_VVPrimeOFXXPrime_audio} shows these local vectors at each point. These vectors and the $\dot{x}_1$ time series were substituted in (\ref{xDot rep}) in order to compute the inner time series, $w_{1}(t)$, for the first unmixed waveform, The result is shown by the thin black line in the left panel of Figure \ref{fig_wWPrimeOfT_audio}.
\item The same procedure was applied to the second unmixed waveform in order to compute its inner time series, $w_{2}(t)$. The result is shown by the thin black line in the right panel of Figure \ref{fig_wWPrimeOfT_audio}.
\item The 500000 samples of the mixed waveform, $x'(t)$, were sorted into a $16 \times 16$ array of bins in $x' \mbox{-space}$, and the distribution of velocities, $\dot{x'}$, in each bin was used to compute the local vectors, $V'_{(i)}(x')$, at each point. These are shown in the right panel of Figure \ref{fig_VVPrimeOFXXPrime_audio}. These vectors and the velocity time series, $\dot{x}'(t)$, were substituted in (\ref{xDot rep}) to compute the inner time series, $w'_{i}(t)$, of the mixed waveforms. These are depicted by the thick gray lines in Figure \ref{fig_wWPrimeOfT_audio}, after they had been multiplied by an overall permutation/reflection matrix.
\end{enumerate}

It is evident that the unmixed and mixed waveforms have inner time series that are almost the same. This demonstrates that an inner time series is not affected by transformations of the measurement time series. In other words, the inner time series encodes sensor-independent information. When each inner time series was played as an audio file, it sounded like a completely intelligible recording of one of the speakers. In each case, the other speaker was not heard, except for a faint buzzing sound in the background. Thus, each inner time series contained all of the semantic information in the unmixed waveform.

Notice that this composite system has an inner time series, $w'_{i}(t)$, which is equal to the collection of the inner time series of its statistically independent subsystems, $w_{1}(t)$ and $w_{2}(t)$. This demonstrates the separability property of the inner time series of a composite system, which was proved in Subsection \ref{composite systems}. Also, notice that the correlation between the time series, $w'_{1}(t)$ and $w'_{2}(t)$, is quite low (-0.0016). As discussed in Subsection \ref{composite systems}, this is expected because these are inner time series of two statistically independent subsystems.

\section{Conclusion}
\label{conclusion}

This paper describes how a time series of sensor measurements can be processed in order to create an inner time series, which is unaffected by the nature of the sensors. Specifically, if a system is observed by two sets of sensors, each set of measurements will lead to the same inner time series if the two sets of measurements are related by any instantaneous, invertible, differentiable transformation. In effect, an inner time series encodes information about the intrinsic nature of the observed system's evolution, without depending on extrinsic factors, such as the observer's choice of sensors. An inner time series is created by statistically processing the local distributions of measurement velocities in order to derive vectors at each point in measurement space. The system's velocity can then be described as a weighted superposition of the local vectors at each point. These time-dependent weights comprise the inner time series. Because they are independent of the coordinate system in measurement space, they represent sensor-independent information about the system's velocity in state space.

The inner time series may be useful in certain practical applications. For instance, it may be used to reduce false negatives in the detection of events of interest. To see this, imagine that the objective is to detect certain ''targeted" movements of a system as it moves through state space, and suppose that this is being done by using a pattern recognition technique to monitor the output of sensors that are observing the system. If the pattern recognition software is trained on the output of calibrated sensors, subsequent sensor drift will cause false negatives to occur. This can be avoided if the pattern recognition algorithm is trained on the inner time series, instead of the time series of raw measurements. As long as the local vectors are computed from recent data from the drifted sensors, the inner time series will be unaffected by sensor drift, and this procedure will sensitively detect the targeted movements. However, it should be noted that this procedure may be accompanied by some false positives. This is because a given inner time series corresponds to multiple measurement time series, which describe trajectories in different regions of the measurement space, as mentioned in Subsection \ref{derivation}.

As an example, consider the output of the moving camera in Subsection \ref{video signal}, and suppose that our objective is to detect camera movements that produce the sensor output shown by the thin black lines in Figure \ref{fig_xXPrimeOfT_video}. Imagine that a pattern recognition algorithm is trained to detect this particular trajectory segment. However, suppose that the camera's lens subsequently ''drifts" so that the targeted camera movements produce the signal shown by the thick gray line in Figure \ref{fig_xXPrimeOfT_video}. In that case, the drifted data will not be recognized, and false negatives will occur.  Now, suppose that the pattern recognition software was trained to recognize the inner time series (Figure \ref{fig_wWPrimeOfT_video}) corresponding to the targeted camera movements. Then, sensor drift will not cause false negatives, as long as the time series to be recognized is processed with local vectors, computed from recently acquired data from the drifted sensors.

As described in Subsection \ref{composite systems}, an inner time series has another attractive property, in addition to its sensor independence. Namely, it automatically provides a separable description of the evolution of a system that is composite in the sense of (\ref{phase space factorization}). To see this, consider the sensors, which observe such a composite system. They may be sensitive to the movements of many subsystems, causing the raw sensor outputs to be unknown, possibly nonlinear, mixtures of many subsystem state variables. Now, suppose that we compute the time series of multi-component weights derived from such mixture measurements.  As proved in Subsection \ref{composite systems}, each component of the inner time series of the composite system is the same as a component of the inner time series of one of its subsystems. In other words, the inner time series of a composite system can be partitioned into groups of components, with each group being equal to the inner time series that would have been derived from a subsystem, if it were possible to observe it alone. Because of this separability property, the inner time series may be useful for detecting a targeted movement of one particular subsystem, in the presence of other independent subsystems. In particular, a pattern recognition procedure can be trained to determine if the components of the inner time series of the targeted movement can be found among the components of the inner time series derived from the mixed measurements of the entire system. An advantage of this procedure is that it is not necessary to use blind source separation (\cite{Comon Jutten}, \cite{Jutten}, \cite{Almeida}, \cite{Levin arXiv}, \cite{Levin LVA-ICA}) to disentangle the measurement time series into its independent components. On the other hand, false positive detections can complicate any such attempt to recognize a targeted signal by its inner time series (instead of its time series of sensor measurements). These errors may occur because multiple different measurement time series may have the same inner time series, as described in Subsection \ref{derivation}.

As an illustrative example, consider the system comprised of two independent audio signals, described in Subsection \ref{two audio signals}, and imagine that our objective is to detect an utterance of the first speaker (left panel of Figure \ref{fig_xOfT_audio}), in the presence of the second speaker (right panel of Figure \ref{fig_xOfT_audio}). It is difficult to determine if this targeted signal is present in the mixtures that are actually measured (Figure \ref{fig_xPrimeOfT_audio}). However, notice that the inner time series of the movement of interest, derived from the unmixed waveforms of a subsystem (the thin black lines in Figure \ref{fig_wWPrimeOfT_audio}), is almost the same as one of the inner time series components, derived from the mixed signals of the composite system (thick gray lines in Figure \ref{fig_wWPrimeOfT_audio}). Therefore, a pattern recognition procedure, which is trained on the inner time series of the unmixed signal, is likely to recognize the targeted signal, even in the presence of signals from other subsystems. 

Some comments on these results:
\begin{enumerate}
\item As stated in Section \ref{introduction}, we have assumed that the sensors produce measurements that are invertibly related to the state variables of the underlying system. This invertibility property can almost be guaranteed by observing the system with a sufficiently large number of independent sensors: specifically, by utilizing at least $2N+1$ independent sensors, where $N$ is the dimension of the system's state space. In this case, the sensors' output lies in an $N \mbox{-dimensional}$ subspace embedded within a space of at least $2N+1$ dimensions. Because an embedding theorem asserts that this subspace is very unlikely to self-intersect (\cite{Sauer}), the points in this subspace are almost certainly invertibly related to the system's state space. Then, dimensional reduction techniques (e.g., \cite{Roweis}) can be used to find the subspace coordinates ($x$) that are invertibly related to the state space points, as desired.  An example was presented in Subsection \ref{video signal}. There, the camera configurations formed a two-dimensional subspace, embedded in a six-dimensional space of raw sensor measurements. This subspace was very unlikely to self-intersect, given that 
$6 > 2N+1 = 5$.  Then, principal components analysis was used to dimensionally reduce the description of each subspace point from six-dimensional coordinates to two-dimensional coordinates ($x$).
\item An inner time series contains information that is intrinsic to the evolution of the observed system, in the sense that it is independent of extrinsic factors, such as the type of sensors used to observe the system. In other words, an inner time series contains information about what is happening ''out there in the real world", independent of how the observer chooses to describe it or experience it. Mathematically speaking, an inner time series is a coordinate-system-independent property of the measurement time series; i.e., its values are the same no matter what measurement coordinate system is used on the system's state space. The local vectors ($V_{(i)}$) also represent a kind of intrinsic structure on state space. These vectors ''mark" state space in a way that is analogous to directional arrows, which mark a physical surface and which can be used as navigational aids, no matter what coordinate system is being used.
\item It is interesting to speculate about the role of inner time series in speech perception. By definition, two people, who understand the same language, tend to perceive the same semantic content of an utterance in that language. Remarkably, this \textit{listener-independence} occurs despite the fact that the listeners may be using significantly different sensors to make measurements of that utterance (e.g., different outer, middle, and inner ears; different cochleas; different neural architectures of the acoustic cortex). This sensor-independence of speech perception suggests that the semantic content of speech may be an inner property; i.e., it may be encoded in the inner time series of speech ($w_{i}(t)$). Specifically, assume that the two listeners have past exposure to statistically similar collections of speech-like sounds. Then, they will perceive the speech-sound manifold to be ``marked" by the same local vectors ($V_{(i)}(x)$), even though they may represent those vectors in different coordinate systems on the speech-sound manifold. Therefore, when the two listeners use (\ref{xDot rep}) to decode an utterance, they will derive the same inner time series, and they will perceive the same semantic content.
\item It is equally remarkable that speech perception is largely \textit{speaker-independent}. Namely, a single listener will instantly recognize that two speakers are uttering the same text. This is true despite the fact that the two sounds were produced by significantly different vocal tracts and may have traversed different regions of the speech-sound manifold. This speaker-independence will occur as long as long as each speaker and the listener have past exposure to statistically similar collections of speech sounds. In that case, because of the above-mentioned listener-independence, each speaker and the listener will derive the same inner time series when they listen to the speaker's utterance. Therefore, if the two speakers have encoded the same semantic content (i.e., the same inner time series) in their utterances, the listener will immediately perceive that they are saying the same thing. Notice that two speakers' utterances, which have the same semantic content, may correspond to two different speech-sound trajectories, which have the same inner time series. Thus, in this speculative scenario, the fact that the same inner time series may be encoded in many measurement time series (see the discussion following (\ref{x rep})) corresponds to the fact that the same semantic content can be expressed by many different voices.
\end{enumerate}

\bibliographystyle{IEEEtran}

\end{document}